# Model-based Cross-correlation Search for Gravitational Waves from the Low-mass X-Ray Binary Scorpius X-1 in LIGO O3 Data

R. Abbott[1], H. Abe[2], F. Acernese[3,4], K. Ackley[5], S. Adhicary[6], N. Adhikari[7], R. X. Adhikari[1], V. K. Adkins[8], V. B. Adya[9], C. Affeldt[10,11], D. Agarwal[12], M. Agathos[13,14], O. D. Aguiar[15], L. Aiello[16], A. Ain[17], P. Ajith[18], T. Akutsu[19,20], S. Albanesi[21,22], R. A. Alfaidi[23], C. Alléné[24], A. Allocca[4,25], P. A. Altin[9], A. Amato[26,27], S. Anand[1], A. Ananyeva[1], S. B. Anderson[1], W. G. Anderson[1], M. Ando[28,29], T. Andrade[30], N. Andres[24], M. Andrés-Carcasona[31], T. Andrić[32], S. Ansoldi[33,34], J. M. Antelis[35], S. Antier[36,37], T. Apostolatos[38], E. Z. Appavuravther[39,40], S. Appert[1], S. K. Apple[41], K. Arai[1], A. Araya[42], M. C. Araya[1], J. S. Areeda[43], M. Arène[44], N. Aritomi[19], N. Arnaud[45,46], M. Arogeti[47], S. M. Aronson[8], H. Asada[48], Y. Aso[50,51], M. Assiduo[52,53], S. Assis de Souza Melo[46], S. M. Aston[54], P. Astone[55], F. Aubin[53], K. AultONeal[35], S. Babak[44], F. Badaracco[56], C. Badger[57], S. Bae[58], Y. Bae[59], S. Bagnasco[22], Y. Bai[1], J. G. Baier[60], J. Baird[44], R. Bajpai[61], T. Baka[62], M. Ball[63], G. Ballardin[46], S. W. Ballmer[64], G. Baltus[65], S. Banagiri[66], B. Banerjee[32], D. Bankar[12], J. C. Barayoga[1], B. C. Barish[1], D. Barker[67], P. Barneo[30], F. Barone[4,68], B. Barr[23], L. Barsotti[69], M. Barsuglia[44], D. Barta[70], J. Bartlett[67], M. A. Barton[23], I. Bartos[71], S. Basak[18], R. Bassiri[72], A. Basti[17,73], M. Bawaj[39,74], J. C. Bayley[23], M. Bazzan[75,76], B. Bécsy[77], V. M. Bedakihale[78], F. Beirnaert[79], M. Bejger[80], I. Belahcene[45], A. S. Bell[23], V. Benedetto[81], D. Beniwal[82], W. Benoit[83], J. D. Bentley[84], M. BenYaala[85], S. Bera[86], M. Berbel[87], F. Bergamin[10,11], B. K. Berger[72], S. Bernuzzi[14], M. Beroiz[1], D. Bersanetti[88], A. Bertolini[27], J. Betzwieser[54], D. Beveridge[89], R. Bhandare[90], A. V. Bhandari[12], U. Bhardwaj[27,37], R. Bhatt[1], D. Bhattacharjee[60,91], S. Bhaumik[71], A. Bianchi[27,92], I. A. Bilenko[93], M. Bilicki[94], G. Billingsley[1], S. Bini[95,96], O. Birnholtz[97], S. Biscans[1,69], M. Bischi[52,53], S. Biscoveanu[69], A. Bisht[10,11], B. Biswas[12], M. Bitossi[17,46], M.-A. Bizouard[36], J. K. Blackburn[1], C. D. Blair[54,89], D. G. Blair[89], R. M. Blair[67], F. Bobba[98,99], N. Bode[10,11], M. Boër[36], G. Bogaert[36], M. Boldrini[55,100], G. N. Bolingbroke[82], L. D. Bonavena[75], R. Bondarescu[30], F. Bondu[101], E. Bonilla[72], R. Bonnand[24], P. Booker[10,11], R. Bork[1], V. Boschi[17], N. Bose[102], S. Bose[12], V. Bossilkov[89], V. Boudart[65], Y. Bouffanais[75,76], A. Bozzi[46], C. Bradaschia[17], P. R. Brady[7,298], A. Bramley[54], A. Branch[54], M. Branchesi[32,103], J. E. Brau[63], M. Breschi[14], T. Briant[104], J. H. Briggs[23], A. Brillet[36], M. Brinkmann[10,11], P. Brockill[7], A. F. Brooks[1], J. Brooks[46], D. D. Brown[82], S. Brunett[1], G. Bruno[56], R. Bruntz[105], J. Bryant[106], F. Bucci[53], J. Buchanan[105], T. Bulik[107], H. J. Bulten[27], A. Buonanno[108,109], K. Burtnyk[67], R. Buscicchio[106,110,111], D. Buskulic[24], C. Buy[112], R. L. Byer[72], G. S. Cabourn Davies[113], G. Cabras[33,34], R. Cabrita[56], L. Cadonati[47], G. Cagnoli[114], C. Cahillane[67], J. Calderón Bustillo[115], J. D. Callaghan[23], T. A. Callister[116,117], E. Calloni[4,25], J. B. Camp[118], M. Canepa[88,119], G. Caneva[31], M. Cannavacciuolo[98], K. C. Cannon[29], H. Cao[82], Z. Cao[120], L. A. Capistran[121], E. Capocasa[19,44], E. Capote[64], G. Carapella[98,99], F. Carbognani[46], M. Carlassara[10,11], J. B. Carlin[122], M. Carpinelli[46,123,124], G. Carrillo[63], J. J. Carter[10,11], G. Carullo[17,73], J. Casanueva Diaz[46], C. Casentini[125,126], G. Castaldi[127], S. Caudill[27,62], M. Cavaglià[91], F. Cavalier[45], R. Cavalieri[46], G. Cella[17], P. Cerdá-Durán[128], E. Cesarini[126], W. Chaibi[36], W. Chakalis[116,117], S. Chalathadka Subrahmanya[84], E. Champion[129], C.-H. Chan[130], C. Chan[29], C. L. Chan[131], K. Chan[131], M. Chan[132], K. Chandra[102], I. P. Chang[130], W. Chang[130], P. Chanial[44,46], S. Chao[130], C. Chapman-Bird[23], P. Charlton[133], E. Chassande-Mottin[44], C. Chatterjee[89], Debarati Chatterjee[12], Deep Chatterjee[7], M. Chaturvedi[90], S. Chaty[44], C. Chen[130,134], D. Chen[50], H. Y. Chen[69], J. Chen[69], K. Chen[135], X. Chen[89], Y.-B. Chen[136], Y.-R. Chen[130], Y. Chen[136], H. Cheng[71], P. Chessa[17,73], H. Y. Cheung[131], H. Y. Chia[71], F. Chiadini[99,137], C-Y. Chiang[138], G. Chiarini[76], R. Chierici[139], A. Chincarini[88], M. L. Chiofalo[17,73], A. Chiummo[46], R. K. Choudhary[89], S. Choudhary[12], N. Christensen[36], Q. Chu[89], Y-K. Chu[138], S. S. Y. Chua[9], K. W. Chung[57], G. Ciani[75,76], P. Ciecielag[80], M. Cieślar[80], M. Cifaldi[125,126], A. A. Ciobanu[82], R. Ciolfi[76,140], F. Clara[67], J. A. Clark[1], T. A. Clarke[5], P. Clearwater[141], S. Clesse[142], F. Cleva[36], E. Coccia[32,103], E. Codazzo[32], P.-F. Cohadon[104], D. E. Cohen[45], M. Colleoni[86], C. G. Collette[143], A. Colombo[110,111], M. Colpi[110,111], C. M. Compton[67], L. Conti[76], S. J. Cooper[106], P. Corban[54], T. R. Corbitt[8], I. Cordero-Carrión[144], S. Corezzi[39,74], N. J. Cornish[77], A. Corsi[145], S. Cortese[46], A. C. Coschizza[146], R. Cotesta[109], R. Cottingham[54], M. W. Coughlin[83], J.-P. Coulon[36], S. T. Countryman[147], B. Cousins[6], P. Couvares[1], D. M. Coward[89], M. J. Cowart[54], D. C. Coyne[1], R. Coyne[148], K. Craig[85], J. D. E. Creighton[7], T. D. Creighton[149], A. W. Criswell[83], M. Croquette[104], S. G. Crowder[150], J. R. Cudell[65], T. J. Cullen[8], A. Cumming[23], R. Cummings[23], E. Cuoco[17,46,151], M. Curyło[107], P. Dabadie[114], T. Dal Canton[45], S. Dall'Osso[55], G. Dálya[79,152], A. Dana[72], B. D'Angelo[88,119], S. Danilishin[26,27], S. D'Antonio[126], K. Danzmann[10,11], C. Darsow-Fromm[84], A. Dasgupta[78], L. E. Datrier[23], Sayantani Datta[153], V. Dattilo[46], I. Dave[90], M. Davier[45], D. Davis[1], M. C. Davis[154], E. J. Daw[155], M. Dax[109], D. De Bra[72,299], M. Deenadayalan[12], J. Degallaix[156], M. De Laurentis[4,25], S. Deléglise[104], V. Del Favero[129], F. De Lillo[56], N. De Lillo[23], D. Dell'Aquila[123,124], W. Del Pozzo[17,73], F. De Matteis[125,126], V. D'Emilio[16], N. Demos[69], T. Dent[115], A. Depasse[56], R. De Pietri[157,158], R. De Rosa[4,25], C. De Rossi[46], R. DeSalvo[127,159], R. De Simone[137], S. Dhurandhar[12], R. Diab[71], M. C. Díaz[149], N. A. Didio[64], T. Dietrich[109], L. Di Fiore[4], C. Di Fronzo[106], C. Di Giorgio[98,99], F. Di Giovanni[128], M. Di Giovanni[32], T. Di Girolamo[4,25], D. Diksha[26,27], A. Di Lieto[17,73], A. Di Michele[74], S. Di Pace[55,100], I. Di Palma[55,100],

F. Di Renzo[17,73], A. K. Divakarla[71], A. Dmitriev[106], Z. Doctor[66], P. P. Doleva[105], L. Donahue[160], L. D'Onofrio[4,25],
F. Donovan[69], K. L. Dooley[16], T. Dooney[62], S. Doravari[12], O. Dorosh[161], M. Drago[55,100], J. C. Driggers[67], Y. Drori[1],
J.-G. Ducoin[44,162], L. Dunn[122], O. Dupletsa[32], O. Durante[98,99], D. D'Urso[123,124], P.-A. Duverne[45], S. E. Dwyer[67], C. Eassa[67],
P. J. Easter[5], M. Ebersold[163], T. Eckhardt[84], G. Eddolls[23], B. Edelman[63], T. B. Edo[1], O. Edy[113], A. Effler[54],
S. Eguchi[132], J. Eichholz[9], S. S. Eikenberry[71], M. Eisenmann[19,24], R. A. Eisenstein[69], A. Ejlli[16], E. Engelby[43],
Y. Enomoto[28], L. Errico[4,25], R. C. Essick[164], H. Estellés[86], D. Estevez[165], T. Etzel[1], M. Evans[69], T. M. Evans[54],
T. Evstafyeva[13], B. E. Ewing[6], F. Fabrizi[52,53], F. Faedi[53], V. Fafone[32,125,126], H. Fair[64], S. Fairhurst[16], P. C. Fan[160],
A. M. Farah[166], B. Farr[63], W. M. Farr[116,117], G. Favaro[75], M. Favata[167], M. Fays[65], M. Fazio[168], J. Feicht[1],
M. M. Fejer[72], E. Fenyvesi[70,169], D. L. Ferguson[170], A. Fernandez-Galiana[69], I. Ferrante[17,73], T. A. Ferreira[15],
F. Fidecaro[17,73], P. Figura[107], A. Fiori[17,73], I. Fiori[46], M. Fishbach[66], R. P. Fisher[105], R. Fittipaldi[99,171],
V. Fiumara[99,172], R. Flaminio[19,24], E. Floden[83], H. K. Fong[29], J. A. Font[128,173], B. Fornal[159], P. W. F. Forsyth[9], A. Franke[84],
S. Frasca[55,100], F. Frasconi[17], J. P. Freed[35], Z. Frei[152], A. Freise[27,92], O. Freitas[174], R. Frey[63], P. Fritschel[69],
V. V. Frolov[54], G. G. Fronzé[22], Y. Fujii[175], Y. Fujikawa[176], Y. Fujimoto[177], P. Fulda[71], M. Fyffe[54], H. A. Gabbard[23],
W. E. Gabella[178], B. U. Gadre[62,109], J. R. Gair[109], J. Gais[131], S. Galaudage[5], R. Gamba[14], D. Ganapathy[69], A. Ganguly[12],
D.-F. Gao[179], D. Gao[72], S. G. Gaonkar[12], B. Garaventa[88,119], C. García-Núñez[180], C. García-Quirós[10,11,86], K. A. Gardner[146],
J. Gargiulo[46], F. Garufi[4,25], C. Gasbarra[125,126], B. Gateley[67], V. Gayathri[71], G.-G. Ge[179], G. Gemme[88], A. Gennai[17],
J. George[90], O. Gerberding[84], L. Gergely[181], S. Ghonge[47], Abhirup Ghosh[109], Archisman Ghosh[79], Shaon Ghosh[167],
Shrobana Ghosh[16], Tathagata Ghosh[12], L. Giacoppo[55,100], J. A. Giaime[8,54], K. D. Giardina[54], D. R. Gibson[180], C. Gier[85],
P. Giri[17,73], F. Gissi[81], S. Gkaitatzis[46], J. Glanzer[8], A. E. Gleckl[43], F. G. Godoy[47], P. Godwin[6], E. Goetz[146], R. Goetz[71],
J. Golomb[1], B. Goncharov[32], G. González[8], M. Gosselin[46], R. Gouaty[24], D. W. Gould[9], S. Goyal[18], B. Grace[9],
A. Grado[4,182], V. Graham[23], M. Granata[156], V. Granata[98], S. Gras[69], P. Grassia[1], C. Gray[67], R. Gray[183], G. Greco[39],
A. C. Green[71], R. Green[16], A. M. Gretarsson[35], E. M. Gretarsson[35], D. Griffith[1], W. L. Griffiths[16], H. L. Griggs[47],
G. Grignani[39,74], A. Grimaldi[95,96], S. J. Grimm[32,103], H. Grote[16], S. Grunewald[109], A. S. Gruson[43], D. Guerra[128],
G. M. Guidi[52,53], A. R. Guimaraes[8], H. K. Gulati[78], F. Gulminelli[184], A. M. Gunny[69], H.-K. Guo[159], Y. Guo[27], Anchal Gupta[1],
Anuradha Gupta[185], P. Gupta[27,62], S. K. Gupta[102], J. Gurs[84], R. Gustafson[186], N. Gutierrez[156], F. Guzman[121], S. Ha[187],
I. P. W. Hadiputrawan[135], L. Haegel[44], S. Haino[138], O. Halim[34], E. D. Hall[69], E. Z. Hamilton[163], G. Hammond[23],
W.-B. Han[188], M. Haney[163], J. Hanks[67], C. Hanna[6], M. D. Hannam[16], O. Hannuksela[27,62], H. Hansen[67], J. Hanson[54],
R. Harada[189], T. Harder[36], K. Haris[27,62], J. Harms[32,103], G. M. Harry[41], I. W. Harry[113], D. Hartwig[84], K. Hasegawa[190],
B. Haskell[80], C.-J. Haster[69], J. S. Hathaway[129], K. Hattori[191], K. Haughian[23], H. Hayakawa[192], K. Hayama[132], F. J. Hayes[23],
J. Healy[129], A. Heidmann[104], A. Heidt[10,11], M. C. Heintze[54], J. Heinze[10,11], J. Heinzel[69], H. Heitmann[36],
F. Hellman[193], P. Hello[45], A. F. Helmling-Cornell[63], G. Hemming[46], M. Hendry[23], I. S. Heng[23], E. Hennes[27],
J.-S. Hennig[26,27], M. Hennig[26,27], C. Henshaw[47], A. G. Hernandez[194], F. Hernandez Vivanco[5], M. Heurs[10,11], A. L. Hewitt[195],
S. Higginbotham[16], S. Hild[26,27], P. Hill[85], Y. Himemoto[196], A. S. Hines[121], N. Hirata[19], C. Hirose[176], T-C. Ho[135], S. Hochheim[10,11],
D. Hofman[156], J. N. Hohmann[84], D. G. Holcomb[154], N. A. Holland[27,92], I. J. Hollows[155], Z. J. Holmes[82], K. Holt[54],
D. E. Holz[166], Q. Hong[130], J. Hough[23], S. Hourihane[1], D. Howell[116,117], E. J. Howell[89], C. G. Hoy[16], D. Hoyland[106],
A. Hreibi[10,11], B-H. Hsieh[190], H-F. Hsieh[130], C. Hsiung[134], H-Y. Huang[138], P. Huang[179], Y-C. Huang[130],
Y.-J. Huang[138], Y. Huang[69], M. T. Hübner[5], A. D. Huddart[197], B. Hughey[35], D. C. Y. Hui[198], V. Hui[24], S. Husa[86],
S. H. Huttner[23], R. Huxford[6], T. Huynh-Dinh[54], J. Hyland[23], G. A. Iandolo[26], S. Ide[199], B. Idzkowski[107], A. Iess[17,151],
K. Inayoshi[200], Y. Inoue[135], P. Iosif[201], J. Irwin[23], Ish Gupta[6], M. Isi[116,117], K. Ito[202], Y. Itoh[177,203], B. R. Iyer[18],
V. JaberianHamedan[89], T. Jacqmin[104], P.-E. Jacquet[104], S. J. Jadhav[204], S. P. Jadhav[12], T. Jain[13], A. L. James[16],
A. Z. Jan[170], K. Jani[178], J. Janquart[27,62], K. Janssens[36,205], N. N. Janthalur[204], P. Jaranowski[206], D. Jariwala[71], S. Jarov[146],
R. Jaume[86], A. C. Jenkins[57], K. Jenner[82], C. Jeon[207], W. Jia[69], J. Jiang[71], H.-B. Jin[208,209], G. R. Johns[105], R. Johnston[23],
N. Johny[10,11], A. W. Jones[89], D. I. Jones[210], P. Jones[106], R. Jones[23], P. Joshi[6], L. Ju[89], K. Jung[187], P. Jung[59],
J. Junker[10,11], V. Juste[165], K. Kaihotsu[202], T. Kajita[211], M. Kakizaki[191], C. Kalaghatgi[27,62,212], V. Kalogera[66], B. Kamai[1],
M. Kamiizumi[192], N. Kanda[177,203], S. Kandhasamy[12], G. Kang[213], J. B. Kanner[1], Y. Kao[130], S. J. Kapadia[18],
D. P. Kapasi[9], S. Karat[1], C. Karathanasis[31], S. Karki[91], R. Kashyap[6], M. Kasprzack[1], W. Kastaun[10,11], T. Kato[190],
S. Katsanevas[46,300], E. Katsavounidis[69], W. Katzman[54], T. Kaur[89], K. Kawabe[67], K. Kawaguchi[190], F. Kéfélian[36],
D. Keitel[86], J. S. Key[214], S. Khadka[72], F. Y. Khalili[93], S. Khan[16], T. Khanam[145], E. A. Khazanov[215], N. Khetan[32,103],
M. Khursheed[90], N. Kijbunchoo[9], C. Kim[207], J. C. Kim[216], J. Kim[217], K. Kim[207], P. Kim[218], W. S. Kim[59],
Y.-M. Kim[187], C. Kimball[66], N. Kimura[192], B. King[219], M. Kinley-Hanlon[23], R. Kirchhoff[10,11], J. S. Kissel[67],
S. Klimenko[71], T. Klinger[16], A. M. Knee[146], N. Knust[10,11], Y. Kobayashi[177], P. Koch[10,11], S. M. Koehlenbeck[10,11],
G. Koekoek[26,27], K. Kohri[220], K. Kokeyama[16], S. Koley[32], P. Kolitsidou[16], M. Kolstein[31], V. Kondrashov[1],
A. K. H. Kong[130], A. Kontos[219], M. Korobko[84], R. V. Kossak[10,11], M. Kovalam[89], N. Koyama[176], D. B. Kozak[1],
C. Kozakai[50], L. Kranzhoff[10,11], V. Kringel[10,11], N. V. Krishnendu[10,11], A. Królak[161,221], G. Kuehn[10,11], P. Kuijer[27],
S. Kulkarni[185], A. Kumar[204], Praveen Kumar[115], Prayush Kumar[18], Rahul Kumar[67], Rakesh Kumar[78], J. Kume[29],
K. Kuns[69], Y. Kuromiya[202], S. Kuroyanagi[222,223], S. Kuwahara[189], K. Kwak[187], G. Lacaille[23], P. Lagabbe[24], D. Laghi[112],
E. Lalande[224], M. Lalleman[205], A. Lamberts[36,225], M. Landry[67], B. B. Lane[69], R. N. Lang[69], J. Lange[170], B. Lantz[72],

I. La Rosa[24], A. Lartaux-Vollard[45], P. D. Lasky[5], J. Lawrence[145], M. Laxen[54], A. Lazzarini[1], C. Lazzaro[75,76], P. Leaci[55,100], S. Leavey[10,11], S. LeBohec[159], Y. K. Lecoeuche[146], E. Lee[190], H. M. Lee[226], H. W. Lee[216], K. Lee[218], R. Lee[130], I. N. Legred[1], J. Lehmann[10,11], A. Lemaître[227], M. Lenti[53,228], M. Leonardi[19], E. Leonova[37], N. Leroy[45], N. Letendre[24], C. Levesque[224], Y. Levin[5], J. N. Leviton[186], K. Leyde[44], A. K. Y. Li[1], B. Li[130], K. L. Li[229], P. Li[230], T. G. F. Li[131], X. Li[136], C-Y. Lin[231], E. T. Lin[130], F-K. Lin[138], F-L. Lin[232], H. L. Lin[135], L. C.-C. Lin[229], F. Linde[27,212], S. D. Linker[127,194], T. B. Littenberg[233], G. C. Liu[134], J. Liu[89], X. Liu[7], F. Llamas[149], R. K. L. Lo[1], T. Lo[130], L. T. London[37,69], A. Longo[234], D. Lopez[163], M. Lopez Portilla[62], M. Lorenzini[125,126], V. Loriette[235], M. Lormand[54], G. Losurdo[17,301], T. P. Lott[47], J. D. Lough[10,11], C. O. Lousto[129], G. Lovelace[43], M. J. Lowry[105], J. F. Lucaccioni[60], H. Lück[10,11], D. Lumaca[125,126], A. P. Lundgren[113], Y. Lung[131], L.-W. Luo[138], A. W. Lussier[224], J. E. Lynam[105], M. Ma'arif[135], R. Macas[113], M. MacInnis[69], D. M. Macleod[16], I. A. O. MacMillan[1], A. Macquet[31,36], I. Maga na Hernandez[7], C. Magazzù[17], R. M. Magee[1], R. Maggiore[27,92,106], M. Magnozzi[88,119], S. Mahesh[236], E. Majorana[55,100], C. N. Makarem[1], I. Maksimovic[235], S. Maliakal[1], A. Malik[90], N. Man[36], V. Mandic[83], V. Mangano[55,100], B. R. Mannix[63], G. L. Mansell[64,67,69], G. Mansingh[41], M. Manske[7], M. Mantovani[46], M. Mapelli[75,76], F. Marchesoni[39,40,237], D. Marín Pina[30], F. Marion[24], Z. Mark[136], S. Márka[147], Z. Márka[147], C. Markakis[183], A. S. Markosyan[72], A. Markowitz[1], E. Maros[1], A. Marquina[144], S. Marsat[112], F. Martelli[52,53], I. W. Martin[23], R. M. Martin[167], M. Martinez[31], V. A. Martinez[71], V. Martinez[114], K. Martinovic[57], D. V. Martynov[106], E. J. Marx[69], H. Masalehdan[84], K. Mason[69], A. Masserot[24], M. Masso-Reid[23], S. Mastrogiovanni[36,44], A. Matas[109], M. Mateu-Lucena[86], M. Matiushechkina[10,11], N. Mavalvala[69], J. J. McCann[89], R. McCarthy[67], D. E. McClelland[9], P. K. McClincy[6], S. McCormick[54], L. McCuller[1,69], G. I. McGhee[23], J. McGinn[23], S. C. McGuire[54], C. McIsaac[113], J. McIver[146], A. McLeod[89], T. McRae[9], S. T. McWilliams[236], D. Meacher[7], M. Mehmet[10,11], A. K. Mehta[109], Q. Meijer[62], A. Melatos[122], G. Mendell[67], A. Menendez-Vazquez[31], C. S. Menoni[168], R. A. Mercer[7], L. Mereni[156], K. Merfeld[63], E. L. Merilh[54], J. D. Merritt[63], M. Merzougui[36], C. Messenger[23], C. Messick[69], P. M. Meyers[136], F. Meylahn[10,11], A. Mhaske[12], A. Miani[95,96], H. Miao[238], I. Michaloliakos[71], C. Michel[156], Y. Michimura[28], H. Middleton[122], D. P. Mihaylov[109], A. Miller[194], A. L. Miller[56], B. Miller[27,37], M. Millhouse[122], J. C. Mills[16], E. Milotti[34,239], Y. Minenkov[126], N. Mio[240], Ll. M. Mir[31], M. Miravet-Tenés[128], A. Mishkin[71], C. Mishra[241], T. Mishra[71], T. Mistry[155], A. L. Mitchell[27,92], S. Mitra[12], V. P. Mitrofanov[93], G. Mitselmakher[71], R. Mittleman[69], O. Miyakawa[192], K. Miyo[192], S. Miyoki[192], Geoffrey Mo[69], L. M. Modafferi[86], E. Moguel[60], K. Mogushi[91], S. R. P. Mohapatra[69], S. R. Mohite[7], M. Molina-Ruiz[193], C. Mondal[184], M. Mondin[194], M. Montani[52,53], C. J. Moore[106], J. Moragues[86], D. Moraru[67], F. Morawski[80], A. More[12], S. More[12], C. Moreno[35], G. Moreno[67], Y. Mori[202], S. Morisaki[7], N. Morisue[177], Y. Moriwaki[191], B. Mours[165], C. M. Mow-Lowry[27,92], S. Mozzon[113], F. Muciaccia[55,100], D. Mukherjee[233], Soma Mukherjee[149], Subroto Mukherjee[78], Suvodip Mukherjee[37,164], N. Mukund[10,11], A. Mullavey[54], J. Munch[82], E. A. Mu niz[64], P. G. Murray[23], S. Muusse[82], S. L. Nadji[10,11], K. Nagano[242], A. Nagar[22,243], T. Nagar[5], K. Nakamura[19], H. Nakano[244], M. Nakano[54,190], Y. Nakayama[202], V. Napolano[46], I. Nardecchia[125,126], T. Narikawa[190], H. Narola[62], L. Naticchioni[55], R. K. Nayak[245], B. F. Neil[89], J. Neilson[81,99], A. Nelson[121], T. J. N. Nelson[54], M. Nery[10,11], P. Neubauer[60], A. Neunzert[214], K. Y. Ng[69], S. W. S. Ng[82], C. Nguyen[44,246], P. Nguyen[63], T. Nguyen[69], L. Nguyen Quynh[247], J. Ni[83], W.-T. Ni[130,179,208], S. A. Nichols[8], G. Nieradka[80], T. Nishimoto[190], A. Nishizawa[29], S. Nissanke[27,37], E. Nitoglia[139], W. Niu[6], F. Nocera[46], M. Norman[16], C. North[16], J. Notte[167], J. Novak[246,248,249,250,251], S. Nozaki[191], G. Nurbek[149], L. K. Nuttall[113], Y. Obayashi[190], J. Oberling[67], B. D. O'Brien[71], J. O'Dell[197], E. Oelker[23], M. Oertel[246,248,249,250,251], W. Ogaki[190], G. Oganesyan[32,103], J. J. Oh[59], K. Oh[198], S. H. Oh[59], T. O'Hanlon[54], M. Ohashi[192], T. Ohashi[177], M. Ohkawa[176], F. Ohme[10,11], H. Ohta[29], Y. Okutani[199], R. Oliveri[252], C. Olivetto[248], K. Oohara[190,253], R. Oram[54], B. O'Reilly[54], R. G. Ormiston[83], N. D. Ormsby[105], M. Orselli[39,74], R. O'Shaughnessy[129], E. O'Shea[254], S. Oshino[192], S. Ossokine[109], C. Osthelder[1], S. Otabe[2], D. J. Ottaway[82], H. Overmier[54], A. E. Pace[6], G. Pagano[17,73], R. Pagano[8], G. Pagliaroli[32,103], A. Pai[102], S. A. Pai[90], S. Pal[245], J. R. Palamos[63], O. Palashov[215], C. Palomba[55], K.-C. Pan[130], P. K. Panda[204], P. T. H. Pang[27,62], F. Pannarale[55,100], B. C. Pant[90], F. H. Panther[89], F. Paoletti[17], A. Paoli[46], A. Paolone[55,255], G. Pappas[201], A. Parisi[17,134,151], J. Park[256], W. Parker[54], D. Pascucci[79], A. Pasqualetti[46], R. Passaquieti[17,73], D. Passuello[17], M. Patel[105], N. R. Patel[67], M. Pathak[82], B. Patricelli[17,73], A. S. Patron[8], S. Paul[63], E. Payne[1], M. Pedraza[1], R. Pedurand[99], R. Pegna[17,73], M. Pegoraro[76], A. Pele[54], F. E. Pe na Arellano[192], S. Penano[72], S. Penn[257], A. Perego[95,96], A. Pereira[114], T. Pereira[258], C. J. Perez[67], C. Périgois[140], C. C. Perkins[71], A. Perreca[95,96], S. Perriès[139], J. W. Perry[27,92], D. Pesios[201], J. Petermann[84], H. P. Pfeiffer[109], H. Pham[54], K. A. Pham[83], K. S. Phukon[27,212], H. Phurailatpam[131], O. J. Piccinni[31,55], M. Pichot[36], M. Piendibene[17,73], F. Piergiovanni[52,53], L. Pierini[55,100], G. Pierra[139], V. Pierro[81,99], G. Pillant[46], M. Pillas[45], F. Pilo[17], L. Pinard[156], C. Pineda-Bosque[194], I. M. Pinto[25,81,99,259], M. Pinto[46], B. J. Piotrzkowski[7], K. Piotrzkowski[56], M. Pirello[67], M. D. Pitkin[195], A. Placidi[39,74], E. Placidi[55,100], M. L. Planas[86], W. Plastino[234,260], R. Poggiani[17,73], E. Polini[24], D. Y. T. Pong[131], S. Ponrathnam[12,302], E. K. Porter[44], C. Posnansky[6], R. Poulton[46], J. Powell[141], M. Pracchia[24], T. Pradier[165], A. K. Prajapati[78], K. Prasai[72], R. Prasanna[204], G. Pratten[106], M. Principe[81,99,259], G. A. Prodi[96,261], L. Prokhorov[106], P. Prosposito[125,126], L. Prudenzi[109], A. Puecher[27,62], M. Punturo[39], F. Puosi[17,73], P. Puppo[55], M. Pürrer[109], H. Qi[16], N. Quartey[105], V. Quetschke[149], P. J. Quinonez[35], R. Quitzow-James[91], F. J. Raab[67], G. Raaijmakers[27,37], H. Radkins[67], N. Radulesco[36], P. Raffai[152], S. X. Rail[224], S. Raja[90], C. Rajan[90], K. E. Ramirez[54], T. D. Ramirez[43], A. Ramos-Buades[109], D. Rana[12],

J. Rana[6], P. R. Rangnekar[72], P. Rapagnani[55,100], A. Ray[7], V. Raymond[16], N. Raza[146], M. Razzano[17,73], J. Read[43], T. Regimbau[24], L. Rei[88], S. Reid[85], S. W. Reid[105], M. Reinhard[71], D. H. Reitze[1], P. Relton[16], A. Renzini[1], P. Rettegno[21,22], B. Revenu[44], J. Reyes[167], A. Reza[27], M. Rezac[43], A. S. Rezaei[55,100], F. Ricci[55,100], D. Richards[197], J. W. Richardson[262], L. Richardson[121], K. Riles[186], S. Rinaldi[17,73], C. Robertson[197], N. A. Robertson[1], R. Robie[1], F. Robinet[45], A. Rocchi[126], S. Rodriguez[43], L. Rolland[24], J. G. Rollins[1], M. Romanelli[101], R. Romano[3,4], C. L. Romel[67], A. Romero[31], I. M. Romero-Shaw[5], J. H. Romie[54], S. Ronchini[32,103], T. J. Roocke[82], L. Rosa[4,25], C. A. Rose[7], D. Rosińska[107], M. P. Ross[263], M. Rossello[86], S. Rowan[23], S. J. Rowlinson[106], Santosh Roy[12], Soumen Roy[62], A. Royzman[159], D. Rozza[123,124], P. Ruggi[46], K. Ruiz-Rocha[178], K. Ryan[67], S. Sachdev[7], T. Sadecki[67], J. Sadiq[115], P. Saffarieh[27,92], S. Saha[130], Y. Saito[192], K. Sakai[264], M. Sakellariadou[57], S. Sakon[6], O. S. Salafia[110,111,265], F. Salces-Carcoba[1], L. Salconi[46], M. Saleem[83], F. Salemi[95,96], M. Sallé[27], A. Samajdar[111], E. J. Sanchez[1], J. H. Sanchez[43], L. E. Sanchez[1], N. Sanchis-Gual[128,266], J. R. Sanders[267], A. Sanuy[30], T. R. Saravanan[12], N. Sarin[5], A. Sasli[201], B. Sassolas[156], H. Satari[89], B. S. Sathyaprakash[6,16], O. Sauter[71], R. L. Savage[67], V. Savant[12], T. Sawada[177], H. L. Sawant[12], S. Sayah[156], D. Schaetzl[1], M. Scheel[136], J. Scheuer[66], M. G. Schiworski[82], P. Schmidt[106], S. Schmidt[62], R. Schnabel[84], M. Schneewind[10,11], R. M. S. Schofield[63], A. Schönbeck[84], B. W. Schulte[10,11], B. F. Schutz[10,11,16], E. Schwartz[16], J. Scott[23], S. M. Scott[9], M. Seglar-Arroyo[24], Y. Sekiguchi[268], D. Sellers[54], A. S. Sengupta[269], D. Sentenac[46], E. G. Seo[131], V. Sequino[4,25], A. Sergeev[215], G. Servignat[249], Y. Setyawati[62], T. Shaffer[67], M. S. Shahriar[66], M. A. Shaikh[18], B. Shams[159], L. Shao[200], A. Sharma[32,103], P. Sharma[90], P. Shawhan[108], N. S. Shcheblanov[227], A. Sheela[241], E. Sheridan[178], Y. Shikano[270,271], M. Shikauchi[29], H. Shimizu[272], K. Shimode[192], H. Shinkai[273], T. Shishido[51], A. Shoda[19], D. H. Shoemaker[69], D. M. Shoemaker[170], S. ShyamSundar[90], M. Sieniawska[56], D. Sigg[67], L. Silenzi[39,40], L. P. Singer[118], D. Singh[6], M. K. Singh[18], N. Singh[107], A. Singha[26,27], A. M. Sintes[86], V. Sipala[123,124], V. Skliris[16], B. J. J. Slagmolen[9], T. J. Slaven-Blair[89], J. Smetana[106], J. R. Smith[43], L. Smith[23], R. J. E. Smith[5], J. Soldateschi[53,228,274], S. N. Somala[275], K. Somiya[2], I. Song[130], K. Soni[12], S. Soni[69], V. Sordini[139], F. Sorrentino[88], N. Sorrentino[17,73], R. Soulard[36], T. Souradeep[12,276], V. Spagnuolo[26,27], A. P. Spencer[23], M. Spera[75,76], P. Spinicelli[46], A. K. Srivastava[78], V. Srivastava[64], C. Stachie[36], F. Stachurski[23], D. A. Steer[44], J. Steinlechner[26,27], S. Steinlechner[26,27], N. Stergioulas[201], D. J. Stops[106], K. A. Strain[23], L. C. Strang[122], G. Stratta[55,277], M. D. Strong[8], A. Strunk[67], R. Sturani[258], A. L. Stuver[154], M. Suchenek[80], S. Sudhagar[12], R. Sugimoto[242,278], H. G. Suh[7], A. G. Sullivan[147], T. Z. Summerscales[279], L. Sun[9], S. Sunil[78], A. Sur[80], J. Suresh[29,56], P. J. Sutton[16], Takamasa Suzuki[176], Takanori Suzuki[2], Toshikazu Suzuki[190], B. L. Swinkels[27], A. Syx[165], M. J. Szczepańczyk[71], P. Szewczyk[107], M. Tacca[27], H. Tagoshi[190], S. C. Tait[23], H. Takahashi[280], R. Takahashi[19], S. Takano[28], H. Takeda[28], M. Takeda[177], C. J. Talbot[85], C. Talbot[69], N. Tamanini[112], K. Tanaka[281], Taiki Tanaka[190], Takahiro Tanaka[282], A. J. Tanasijczuk[56], S. Tanioka[192], D. B. Tanner[71], D. Tao[1], L. Tao[71], R. D. Tapia[6], E. N. Tapia San Martín[27], C. Taranto[125], A. Taruya[283], J. D. Tasson[160], R. Tenorio[86], J. E. S. Terhune[154], L. Terkowski[84], H. Themann[194], M. P. Thirugnanasambandam[12], M. Thomas[54], P. Thomas[67], S. Thomas[43], D. Thompson[160], E. E. Thompson[47], J. E. Thompson[16], S. R. Thondapu[90], K. A. Thorne[54], E. Thrane[5], Shubhanshu Tiwari[163], Srishti Tiwari[12], V. Tiwari[16], A. M. Toivonen[83], A. E. Tolley[113], T. Tomaru[19], T. Tomura[192], M. Tonelli[17,73], A. Torres-Forné[128], C. I. Torrie[1], I. Tosta e Melo[124], E. Tournefier[24], D. Töyrä[9], A. Trapananti[39,40], F. Travasso[39,40], G. Traylor[54], J. Trenado[30], M. Trevor[108], M. C. Tringali[46], A. Tripathee[186], L. Troiano[99,284], A. Trovato[34,239], L. Trozzo[4,192], R. J. Trudeau[1], D. Tsai[130], K. W. Tsang[27,62,285], T. Tsang[286], J-S. Tsao[232], M. Tse[69], R. Tso[136], S. Tsuchida[177], L. Tsukada[6], D. Tsuna[29], T. Tsutsui[29], K. Turbang[205,287], M. Turconi[36], C. Turski[79], D. Tuyenbayev[177], H. Ubach[30], A. S. Ubhi[106], N. Uchikata[190], T. Uchiyama[192], R. P. Udall[1], A. Ueda[288], T. Uehara[289,290], K. Ueno[29], G. Ueshima[291], C. S. Unnikrishnan[292], A. L. Urban[8], T. Ushiba[192], A. Utina[26,27], H. Vahlbruch[10,11], N. Vaidya[1], G. Vajente[1], A. Vajpeyi[5], G. Valdes[121], M. Valentini[95,96,185], S. Vallero[22], V. Valsan[7], N. van Bakel[27], M. van Beuzekom[27], M. van Dael[27,293], J. F. J. van den Brand[26,27,92], C. Van Den Broeck[27,62], D. C. Vander-Hyde[64], A. Van de Walle[45], J. van Dongen[27,92], H. van Haevermaet[205], J. V. van Heijningen[56], J. Vanosky[1], M. H. P. M. van Putten[294], Z. van Ranst[26], N. van Remortel[205], M. Vardaro[27,212], A. F. Vargas[122], V. Varma[109], M. Vasúth[70], A. Vecchio[106], G. Vedovato[76], J. Veitch[23], P. J. Veitch[82], J. Venneberg[10,11], G. Venugopalan[1], P. Verdier[139], D. Verkindt[24], P. Verma[161], Y. Verma[90], S. M. Vermeulen[16], D. Veske[147], F. Vetrano[52], A. Viceré[52,53], S. Vidyant[64], A. D. Viets[295], A. Vijaykumar[18], V. Villa-Ortega[115], J.-Y. Vinet[36], A. Virtuoso[34,239], S. Vitale[69], H. Vocca[39,74], E. R. G. von Reis[67], J. S. A. von Wrangel[10,11], C. Vorvick[67], S. P. Vyatchanin[93], L. E. Wade[60], M. Wade[60], K. J. Wagner[129], R. C. Walet[27], M. Walker[105], G. S. Wallace[85], L. Wallace[1], J. Wang[179], J. Z. Wang[186], W. H. Wang[149], R. L. Ward[9], J. Warner[67], M. Was[24], T. Washimi[19], N. Y. Washington[1], K. Watada[105], D. Watarai[189], J. Watchi[143], K. E. Wayt[60], B. Weaver[67], C. R. Weaving[113], S. A. Webster[23], M. Weinert[10,11], A. J. Weinstein[1], R. Weiss[69], C. M. Weller[263], R. A. Weller[178], F. Wellmann[10,11], L. Wen[89], P. Weßels[10,11], K. Wette[9], J. T. Whelan[129], D. D. White[43], B. F. Whiting[71], C. Whittle[69], O. S. Wilk[60], D. Wilken[10,11,11], C. E. Williams[160], D. Williams[23], M. J. Williams[23], A. R. Williamson[113], J. L. Willis[1], B. Willke[10,11], C. C. Wipf[1], G. Woan[23], J. Woehler[10,11], J. K. Wofford[129], I. A. Wojtowicz[160], D. Wong[146], I. C. F. Wong[131], M. Wright[23], C. Wu[130], D. S. Wu[10,11], H. Wu[130], D. M. Wysocki[7], L. Xiao[1], N. Yadav[80], T. Yamada[272], H. Yamamoto[1], K. Yamamoto[191], T. Yamamoto[192], K. Yamashita[202], R. Yamazaki[199], F. W. Yang[159], K. Z. Yang[83], L. Yang[168], Y.-C. Yang[130], Y. Yang[296], Yang Yang[71], M. J. Yap[9], D. W. Yeeles[16], S.-W. Yeh[130],

A. B. Yelikar[129], J. Yokoyama[28,29,303], T. Yokozawa[192], J. Yoo[254], T. Yoshioka[202], Hang Yu[136], Haocun Yu[69], H. Yuzurihara[190], A. Zadrożny[161], M. Zanolin[35], S. Zeidler[297], T. Zelenova[46], J.-P. Zendri[76], M. Zevin[166], M. Zhan[179], H. Zhang[232], J. Zhang[9], L. Zhang[1], R. Zhang[71], T. Zhang[106], Y. Zhang[121], C. Zhao[89], G. Zhao[143], Y. Zhao[19,190], Yue Zhao[159], Y. Zheng[91], R. Zhou[193], X. J. Zhu[5], Z.-H. Zhu[120,230], A. B. Zimmerman[170], M. E. Zucker[1,69], and J. Zweizig[1]

The LIGO Scientific Collaboration, the Virgo Collaboration, and the KAGRA Collaboration

[1] LIGO Laboratory, California Institute of Technology, Pasadena, CA 91125, USA
[2] Graduate School of Science, Tokyo Institute of Technology, Meguro-ku, Tokyo 152-8551, Japan
[3] Dipartimento di Farmacia, Università di Salerno, I-84084 Fisciano, Salerno, Italy
[4] INFN, Sezione di Napoli, I-80126 Napoli, Italy
[5] OzGrav, School of Physics & Astronomy, Monash University, Clayton, VIC 3800, Australia
[6] The Pennsylvania State University, University Park, PA 16802, USA
[7] University of Wisconsin-Milwaukee, Milwaukee, WI 53201, USA
[8] Louisiana State University, Baton Rouge, LA 70803, USA
[9] OzGrav, Australian National University, Canberra, Australian Capital Territory 0200, Australia
[10] Max Planck Institute for Gravitational Physics (Albert Einstein Institute), D-30167 Hannover, Germany
[11] Leibniz Universität Hannover, D-30167 Hannover, Germany
[12] Inter-University Centre for Astronomy and Astrophysics, Pune 411007, India
[13] University of Cambridge, Cambridge CB2 1TN, UK
[14] Theoretisch-Physikalisches Institut, Friedrich-Schiller-Universität Jena, D-07743 Jena, Germany
[15] Instituto Nacional de Pesquisas Espaciais, 12227-010 São José dos Campos, São Paulo, Brazil
[16] Cardiff University, Cardiff CF24 3AA, UK
[17] INFN, Sezione di Pisa, I-56127 Pisa, Italy
[18] International Centre for Theoretical Sciences, Tata Institute of Fundamental Research, Bengaluru 560089, India
[19] Gravitational Wave Science Project, National Astronomical Observatory of Japan (NAOJ), Mitaka City, Tokyo 181-8588, Japan
[20] Advanced Technology Center, National Astronomical Observatory of Japan (NAOJ), Mitaka City, Tokyo 181-8588, Japan
[21] Dipartimento di Fisica, Università degli Studi di Torino, I-10125 Torino, Italy
[22] INFN Sezione di Torino, I-10125 Torino, Italy
[23] SUPA, University of Glasgow, Glasgow G12 8QQ, UK
[24] Univ. Savoie Mont Blanc, CNRS, Laboratoire d'Annecy de Physique des Particules—IN2P3, F-74000 Annecy, France
[25] Università di Napoli "Federico II," I-80126 Napoli, Italy
[26] Maastricht University, 6200 MD Maastricht, The Netherlands
[27] Nikhef, 1098 XG Amsterdam, The Netherlands
[28] Department of Physics, The University of Tokyo, Bunkyo-ku, Tokyo 113-0033, Japan
[29] Research Center for the Early Universe (RESCEU), The University of Tokyo, Bunkyo-ku, Tokyo 113-0033, Japan
[30] Institut de Ciències del Cosmos (ICCUB), Universitat de Barcelona, Barcelona, E-08028, Spain
[31] Institut de Física d'Altes Energies (IFAE), Barcelona Institute of Science and Technology, and ICREA, E-08193 Barcelona, Spain
[32] Gran Sasso Science Institute (GSSI), I-67100 L'Aquila, Italy
[33] Dipartimento di Scienze Matematiche, Informatiche e Fisiche, Università di Udine, I-33100 Udine, Italy
[34] INFN, Sezione di Trieste, I-34127 Trieste, Italy
[35] Embry-Riddle Aeronautical University, Prescott, AZ 86301, USA
[36] Artemis, Université Côte d'Azur, Observatoire de la Côte d'Azur, CNRS, F-06304 Nice, France
[37] GRAPPA, Anton Pannekoek Institute for Astronomy and Institute for High-Energy Physics, University of Amsterdam, 1098 XH Amsterdam, The Netherlands
[38] Department of Physics, National and Kapodistrian University of Athens, 15771 Ilissia, Greece
[39] INFN, Sezione di Perugia, I-06123 Perugia, Italy
[40] Università di Camerino, Dipartimento di Fisica, I-62032 Camerino, Italy
[41] American University, Washington, DC 20016, USA
[42] Earthquake Research Institute, The University of Tokyo, Bunkyo-ku, Tokyo 113-0032, Japan
[43] California State University Fullerton, Fullerton, CA 92831, USA
[44] Université de Paris, CNRS, Astroparticule et Cosmologie, F-75006 Paris, France
[45] Université Paris-Saclay, CNRS/IN2P3, IJCLab, F-91405 Orsay, France
[46] European Gravitational Observatory (EGO), I-56021 Cascina, Pisa, Italy
[47] Georgia Institute of Technology, Atlanta, GA 30332, USA
[48] Department of Mathematics and Physics, Graduate School of Science and Technology, Hirosaki University, Hirosaki, Aomori 036-8561, Japan
[49] Royal Holloway, University of London, London TW20 0EX, UK
[50] Kamioka Branch, National Astronomical Observatory of Japan (NAOJ), Kamioka-cho, Hida City, Gifu 506-1205, Japan
[51] The Graduate University for Advanced Studies (SOKENDAI), Mitaka City, Tokyo 181-8588, Japan
[52] Università degli Studi di Urbino "Carlo Bo," I-61029 Urbino, Italy
[53] INFN, Sezione di Firenze, I-50019 Sesto Fiorentino, Firenze, Italy
[54] LIGO Livingston Observatory, Livingston, LA 70754, USA
[55] INFN, Sezione di Roma, I-00185 Roma, Italy
[56] Université catholique de Louvain, B-1348 Louvain-la-Neuve, Belgium
[57] King's College London, University of London, London WC2R 2LS, UK
[58] Korea Institute of Science and Technology Information, Daejeon 34141, Republic of Korea
[59] National Institute for Mathematical Sciences, Daejeon 34047, Republic of Korea
[60] Kenyon College, Gambier, OH 43022, USA
[61] School of High Energy Accelerator Science, The Graduate University for Advanced Studies (SOKENDAI), Tsukuba City, Ibaraki 305-0801, Japan
[62] Institute for Gravitational and Subatomic Physics (GRASP), Utrecht University, 3584 CC Utrecht, The Netherlands
[63] University of Oregon, Eugene, OR 97403, USA
[64] Syracuse University, Syracuse, NY 13244, USA
[65] Université de Liège, B-4000 Liège, Belgium
[66] Northwestern University, Evanston, IL 60208, USA
[67] LIGO Hanford Observatory, Richland, WA 99352, USA
[68] Dipartimento di Medicina, Chirurgia e Odontoiatria "Scuola Medica Salernitana," Università di Salerno, I-84081 Baronissi, Salerno, Italy

[69] LIGO Laboratory, Massachusetts Institute of Technology, Cambridge, MA 02139, USA
[70] Wigner RCP, RMKI, H-1121 Budapest, Hungary
[71] University of Florida, Gainesville, FL 32611, USA
[72] Stanford University, Stanford, CA 94305, USA
[73] Università di Pisa, I-56127 Pisa, Italy
[74] Università di Perugia, I-06123 Perugia, Italy
[75] Università di Padova, Dipartimento di Fisica e Astronomia, I-35131 Padova, Italy
[76] INFN, Sezione di Padova, I-35131 Padova, Italy
[77] Montana State University, Bozeman, MT 59717, USA
[78] Institute for Plasma Research, Bhat, Gandhinagar 382428, India
[79] Universiteit Gent, B-9000 Gent, Belgium
[80] Nicolaus Copernicus Astronomical Center, Polish Academy of Sciences, 00-716, Warsaw, Poland
[81] Dipartimento di Ingegneria, Università del Sannio, I-82100 Benevento, Italy
[82] OzGrav, University of Adelaide, Adelaide, South Australia 5005, Australia
[83] University of Minnesota, Minneapolis, MN 55455, USA
[84] Universität Hamburg, D-22761 Hamburg, Germany
[85] SUPA, University of Strathclyde, Glasgow G1 1XQ, UK
[86] IAC3–IEEC, Universitat de les Illes Balears, E-07122 Palma de Mallorca, Spain
[87] Departamento de Matemáticas, Universitat Autònoma de Barcelona, E-08193 Bellaterra (Barcelona), Spain
[88] INFN, Sezione di Genova, I-16146 Genova, Italy
[89] OzGrav, University of Western Australia, Crawley, Western Australia 6009, Australia
[90] RRCAT, Indore, Madhya Pradesh 452013, India
[91] Missouri University of Science and Technology, Rolla, MO 65409, USA
[92] Department of Physics and Astronomy, Vrije Universiteit Amsterdam, 1081 HV Amsterdam, The Netherlands
[93] Lomonosov Moscow State University, Moscow 119991, Russia
[94] Center for Theoretical Physics, Polish Academy of Sciences, 02-668, Warsaw, Poland
[95] Università di Trento, Dipartimento di Fisica, I-38123 Povo, Trento, Italy
[96] INFN, Trento Institute for Fundamental Physics and Applications, I-38123 Povo, Trento, Italy
[97] Bar-Ilan University, Ramat Gan, 5290002, Israel
[98] Dipartimento di Fisica "E.R. Caianiello," Università di Salerno, I-84084 Fisciano, Salerno, Italy
[99] INFN, Sezione di Napoli, Gruppo Collegato di Salerno, I-80126 Napoli, Italy
[100] Università di Roma "La Sapienza," I-00185 Roma, Italy
[101] Univ Rennes, CNRS, Institut FOTON—UMR 6082, F-3500 Rennes, France
[102] Indian Institute of Technology Bombay, Powai, Mumbai 400 076, India
[103] INFN, Laboratori Nazionali del Gran Sasso, I-67100 Assergi, Italy
[104] Laboratoire Kastler Brossel, Sorbonne Université, CNRS, ENS-Université PSL, Collège de France, F-75005 Paris, France
[105] Christopher Newport University, Newport News, VA 23606, USA
[106] University of Birmingham, Birmingham B15 2TT, UK
[107] Astronomical Observatory Warsaw University, 00-478 Warsaw, Poland
[108] University of Maryland, College Park, MD 20742, USA
[109] Max Planck Institute for Gravitational Physics (Albert Einstein Institute), D-14476 Potsdam, Germany
[110] Università degli Studi di Milano-Bicocca, I-20126 Milano, Italy
[111] INFN, Sezione di Milano-Bicocca, I-20126 Milano, Italy
[112] L2IT, Laboratoire des 2 Infinis—Toulouse, Université de Toulouse, CNRS/IN2P3, UPS, F-31062 Toulouse Cedex 9, France
[113] University of Portsmouth, Portsmouth, PO1 3FX, UK
[114] Université de Lyon, Université Claude Bernard Lyon 1, CNRS, Institut Lumière Matière, F-69622 Villeurbanne, France
[115] IGFAE, Universidade de Santiago de Compostela, E-15782 Spain
[116] Stony Brook University, Stony Brook, NY 11794, USA
[117] Center for Computational Astrophysics, Flatiron Institute, New York, NY 10010, USA
[118] NASA Goddard Space Flight Center, Greenbelt, MD 20771, USA
[119] Dipartimento di Fisica, Università degli Studi di Genova, I-16146 Genova, Italy
[120] Department of Astronomy, Beijing Normal University, Beijing 100875, People's Republic of China
[121] Texas A&M University, College Station, TX 77843, USA
[122] OzGrav, University of Melbourne, Parkville, VIC 3010, Australia
[123] Università degli Studi di Sassari, I-07100 Sassari, Italy
[124] INFN, Laboratori Nazionali del Sud, I-95125 Catania, Italy
[125] Università di Roma Tor Vergata, I-00133 Roma, Italy
[126] INFN, Sezione di Roma Tor Vergata, I-00133 Roma, Italy
[127] University of Sannio at Benevento, I-82100 Benevento, Italy and INFN, Sezione di Napoli, I-80100 Napoli, Italy
[128] Departamento de Astronomía y Astrofísica, Universitat de València, E-46100 Burjassot, València, Spain
[129] Rochester Institute of Technology, Rochester, NY 14623, USA
[130] National Tsing Hua University, Hsinchu City, 30013, Taiwan
[131] The Chinese University of Hong Kong, Shatin, NT, Hong Kong
[132] Department of Applied Physics, Fukuoka University, Jonan, Fukuoka City, Fukuoka 814-0180, Japan
[133] OzGrav, Charles Sturt University, Wagga Wagga, New South Wales 2678, Australia
[134] Department of Physics, Tamkang University, Danshui Dist., New Taipei City 25137, Taiwan
[135] Department of Physics, Center for High Energy and High Field Physics, National Central University, Zhongli District, Taoyuan City 32001, Taiwan
[136] CaRT, California Institute of Technology, Pasadena, CA 91125, USA
[137] Dipartimento di Ingegneria Industriale (DIIN), Università di Salerno, I-84084 Fisciano, Salerno, Italy
[138] Institute of Physics, Academia Sinica, Nankang, Taipei 11529, Taiwan
[139] Université Lyon, Université Claude Bernard Lyon 1, CNRS, IP2I Lyon / IN2P3, UMR 5822, F-69622 Villeurbanne, France
[140] INAF, Osservatorio Astronomico di Padova, I-35122 Padova, Italy
[141] OzGrav, Swinburne University of Technology, Hawthorn, VIC 3122, Australia
[142] Université libre de Bruxelles, B-1050 Bruxelles, Belgium
[143] Université Libre de Bruxelles, B-1050 Brussels, Belgium
[144] Departamento de Matemáticas, Universitat de València, E-46100 Burjassot, València, Spain

[145] Texas Tech University, Lubbock, TX 79409, USA
[146] University of British Columbia, Vancouver, BC V6T 1Z4, Canada
[147] Columbia University, New York, NY 10027, USA
[148] University of Rhode Island, Kingston, RI 02881, USA
[149] The University of Texas Rio Grande Valley, Brownsville, TX 78520, USA
[150] Bellevue College, Bellevue, WA 98007, USA
[151] Scuola Normale Superiore, I-56126 Pisa, Italy
[152] Eötvös University, Budapest 1117, Hungary
[153] Chennai Mathematical Institute, Chennai 603103, India
[154] Villanova University, Villanova, PA 19085, USA
[155] The University of Sheffield, Sheffield S10 2TN, UK
[156] Université Lyon, Université Claude Bernard Lyon 1, CNRS, Laboratoire des Matériaux Avancés (LMA), IP2I Lyon / IN2P3, UMR 5822, F-69622 Villeurbanne, France
[157] Dipartimento di Scienze Matematiche, Fisiche e Informatiche, Università di Parma, I-43124 Parma, Italy
[158] INFN, Sezione di Milano Bicocca, Gruppo Collegato di Parma, I-43124 Parma, Italy
[159] The University of Utah, Salt Lake City, UT 84112, USA
[160] Carleton College, Northfield, MN 55057, USA
[161] National Center for Nuclear Research, 05-400 Świerk-Otwock, Poland
[162] Institut d'Astrophysique de Paris, Sorbonne Université, CNRS, UMR 7095, F-75014 Paris, France
[163] University of Zurich, Winterthurerstrasse 190, 8057 Zurich, Switzerland
[164] Perimeter Institute, Waterloo, ON N2L 2Y5, Canada
[165] Université de Strasbourg, CNRS, IPHC UMR 7178, F-67000 Strasbourg, France
[166] University of Chicago, Chicago, IL 60637, USA
[167] Montclair State University, Montclair, NJ 07043, USA
[168] Colorado State University, Fort Collins, CO 80523, USA
[169] Institute for Nuclear Research, H-4026 Debrecen, Hungary
[170] University of Texas, Austin, TX 78712, USA
[171] CNR-SPIN, I-84084 Fisciano, Salerno, Italy
[172] Scuola di Ingegneria, Università della Basilicata, I-85100 Potenza, Italy
[173] Observatori Astronòmic, Universitat de València, E-46980 Paterna, València, Spain
[174] Centro de Física das Universidades do Minho e do Porto, Universidade do Minho, PT-4710-057 Braga, Portugal
[175] Department of Astronomy, The University of Tokyo, Mitaka City, Tokyo 181-8588, Japan
[176] Faculty of Engineering, Niigata University, Nishi-ku, Niigata City, Niigata 950-2181, Japan
[177] Department of Physics, Graduate School of Science, Osaka City University, Sumiyoshi-ku, Osaka City, Osaka 558-8585, Japan
[178] Vanderbilt University, Nashville, TN 37235, USA
[179] State Key Laboratory of Magnetic Resonance and Atomic and Molecular Physics, Innovation Academy for Precision Measurement Science and Technology (APM), Chinese Academy of Sciences, Xiao Hong Shan, Wuhan 430071, People's Republic of China
[180] SUPA, University of the West of Scotland, Paisley PA1 2BE, UK
[181] University of Szeged, Dóm tér 9, Szeged 6720, Hungary
[182] INAF, Osservatorio Astronomico di Capodimonte, I-80131 Napoli, Italy
[183] Queen Mary University of London, London E1 4NS, UK
[184] Université de Normandie, ENSICAEN, UNICAEN, CNRS/IN2P3, LPC Caen, F-14000 Caen, France
[185] The University of Mississippi, University, MS 38677, USA
[186] University of Michigan, Ann Arbor, MI 48109, USA
[187] Ulsan National Institute of Science and Technology, Ulsan 44919, Republic of Korea
[188] Shanghai Astronomical Observatory, Chinese Academy of Sciences, Shanghai 200030, People's Republic of China
[189] University of Tokyo, Tokyo, 113-0033, Japan
[190] Institute for Cosmic Ray Research (ICRR), KAGRA Observatory, The University of Tokyo, Kashiwa City, Chiba 277-8582, Japan
[191] Faculty of Science, University of Toyama, Toyama City, Toyama 930-8555, Japan
[192] Institute for Cosmic Ray Research (ICRR), KAGRA Observatory, The University of Tokyo, Kamioka-cho, Hida City, Gifu 506-1205, Japan
[193] University of California, Berkeley, CA 94720, USA
[194] California State University, Los Angeles, Los Angeles, CA 90032, USA
[195] Lancaster University, Lancaster LA1 4YW, UK
[196] College of Industrial Technology, Nihon University, Narashino City, Chiba 275-8575, Japan
[197] Rutherford Appleton Laboratory, Didcot OX11 0DE, UK
[198] Department of Astronomy & Space Science, Chungnam National University, Yuseong-gu, Daejeon 34134, Republic of Korea
[199] Department of Physical Sciences, Aoyama Gakuin University, Sagamihara City, Kanagawa 252-5258, Japan
[200] Kavli Institute for Astronomy and Astrophysics, Peking University, Haidian District, Beijing 100871, People's Republic of China
[201] Department of Physics, Aristotle University of Thessaloniki, 54124 Thessaloniki, Greece
[202] Graduate School of Science and Engineering, University of Toyama, Toyama City, Toyama 930-8555, Japan
[203] Nambu Yoichiro Institute of Theoretical and Experimental Physics (NITEP), Osaka City University, Sumiyoshi-ku, Osaka City, Osaka 558-8585, Japan
[204] Directorate of Construction, Services & Estate Management, Mumbai 400094, India
[205] Universiteit Antwerpen, B-2000 Antwerpen, Belgium
[206] University of Białystok, 15-424 Białystok, Poland
[207] Ewha Womans University, Seoul 03760, Republic of Korea
[208] National Astronomical Observatories, Chinese Academic of Sciences, Chaoyang District, Beijing, People's Republic of China
[209] School of Astronomy and Space Science, University of Chinese Academy of Sciences, Chaoyang District, Beijing, People's Republic of China
[210] University of Southampton, Southampton SO17 1BJ, UK
[211] Institute for Cosmic Ray Research (ICRR), The University of Tokyo, Kashiwa City, Chiba 277-8582, Japan
[212] Institute for High-Energy Physics, University of Amsterdam, 1098 XH Amsterdam, The Netherlands
[213] Chung-Ang University, Seoul 06974, Republic of Korea
[214] University of Washington Bothell, Bothell, WA 98011, USA
[215] Institute of Applied Physics, Nizhny Novgorod, 603950, Russia
[216] Inje University Gimhae, South Gyeongsang 50834, Republic of Korea
[217] Department of Physics, Myongji University, Yongin 17058, Republic of Korea
[218] Sungkyunkwan University, Seoul 03063, Republic of Korea

[219] Bard College, Annandale-On-Hudson, NY 12504, USA
[220] Institute of Particle and Nuclear Studies (IPNS), High Energy Accelerator Research Organization (KEK), Tsukuba City, Ibaraki 305-0801, Japan
[221] Institute of Mathematics, Polish Academy of Sciences, 00656 Warsaw, Poland
[222] Instituto de Fisica Teorica, E-28049 Madrid, Spain
[223] Department of Physics, Nagoya University, Chikusa-ku, Nagoya, Aichi 464-8602, Japan
[224] Université de Montréal/Polytechnique, Montreal, QC H3T 1J4, Canada
[225] Laboratoire Lagrange, Université Côte d'Azur, Observatoire Côte d'Azur, CNRS, F-06304 Nice, France
[226] Seoul National University, Seoul 08826, Republic of Korea
[227] NAVIER, École des Ponts, Univ Gustave Eiffel, CNRS, Marne-la-Vallée, France
[228] Università di Firenze, Sesto Fiorentino I-50019, Italy
[229] Department of Physics, National Cheng Kung University, Tainan City 701, Taiwan
[230] School of Physics and Technology, Wuhan University, Wuhan, Hubei, 430072, People's Republic of China
[231] National Center for High-performance Computing, National Applied Research Laboratories, Hsinchu Science Park, Hsinchu City 30076, Taiwan
[232] Department of Physics, National Taiwan Normal University, Section 4, Taipei 116, Taiwan
[233] NASA Marshall Space Flight Center, Huntsville, AL 35811, USA
[234] INFN, Sezione di Roma Tre, I-00146 Roma, Italy
[235] ESPCI, CNRS, F-75005 Paris, France
[236] West Virginia University, Morgantown, WV 26506, USA
[237] School of Physics Science and Engineering, Tongji University, Shanghai 200092, People's Republic of China
[238] Tsinghua University, Beijing 100084, People's Republic of China
[239] Dipartimento di Fisica, Università di Trieste, I-34127 Trieste, Italy
[240] Institute for Photon Science and Technology, The University of Tokyo, Bunkyo-ku, Tokyo 113-8656, Japan
[241] Indian Institute of Technology Madras, Chennai 600036, India
[242] Institute of Space and Astronautical Science (JAXA), Chuo-ku, Sagamihara City, Kanagawa 252-0222, Japan
[243] Institut des Hautes Etudes Scientifiques, F-91440 Bures-sur-Yvette, France
[244] Faculty of Law, Ryukoku University, Fushimi-ku, Kyoto City, Kyoto 612-8577, Japan
[245] Indian Institute of Science Education and Research, Kolkata, Mohanpur, West Bengal 741252, India
[246] Université de Paris, F-75006 Paris, France
[247] Department of Physics, University of Notre Dame, Notre Dame, IN 46556, USA
[248] Centre national de la recherche scientifique, F-75016 Paris, France
[249] Laboratoire Univers et Théories, Observatoire de Paris, F-92190 Meudon, France
[250] Observatoire de Paris, F-75014 Paris, France
[251] Université PSL, F-75006 Paris, France
[252] Institute of Physics of the Czech Academy of Sciences, 182 00 Praha 8, Czechia
[253] Graduate School of Science and Technology, Niigata University, Nishi-ku, Niigata City, Niigata 950-2181, Japan
[254] Cornell University, Ithaca, NY 14850, USA
[255] Consiglio Nazionale delle Ricerche—Istituto dei Sistemi Complessi, I-00185 Roma, Italy
[256] Korea Astronomy and Space Science Institute (KASI), Yuseong-gu, Daejeon 34055, Republic of Korea
[257] Hobart and William Smith Colleges, Geneva, NY 14456, USA
[258] International Institute of Physics, Universidade Federal do Rio Grande do Norte, Natal RN 59078-970, Brazil
[259] Museo Storico della Fisica e Centro Studi e Ricerche "Enrico Fermi," I-00184 Roma, Italy
[260] Dipartimento di Matematica e Fisica, Università degli Studi Roma Tre, I-00146 Roma, Italy
[261] Università di Trento, Dipartimento di Matematica, I-38123 Povo, Trento, Italy
[262] University of California, Riverside, Riverside, CA 92521, USA
[263] University of Washington, Seattle, WA 98195, USA
[264] Department of Electronic Control Engineering, National Institute of Technology, Nagaoka College, Nagaoka City, Niigata 940-8532, Japan
[265] INAF, Osservatorio Astronomico di Brera sede di Merate, I-23807 Merate, Lecco, Italy
[266] Departamento de Matemática da Universidade de Aveiro and Centre for Research and Development in Mathematics and Applications, 3810-183 Aveiro, Portugal
[267] Marquette University, Milwaukee, WI 53233, USA
[268] Faculty of Science, Toho University, Funabashi City, Chiba 274-8510, Japan
[269] Indian Institute of Technology, Palaj, Gandhinagar, Gujarat 382355, India
[270] Graduate School of Science and Technology, Gunma University, Maebashi, Gunma 371-8510, Japan
[271] Institute for Quantum Studies, Chapman University, Orange, CA 92866, USA
[272] Accelerator Laboratory, High Energy Accelerator Research Organization (KEK), Tsukuba City, Ibaraki 305-0801, Japan
[273] Faculty of Information Science and Technology, Osaka Institute of Technology, Hirakata City, Osaka 573-0196, Japan
[274] INAF, Osservatorio Astrofisico di Arcetri, I-50125 Firenze, Italy
[275] Indian Institute of Technology Hyderabad, Sangareddy, Khandi, Telangana 502285, India
[276] Indian Institute of Science Education and Research, Pune, Maharashtra 411008, India
[277] Istituto di Astrofisica e Planetologia Spaziali di Roma, I-00133 Roma, Italy
[278] Department of Space and Astronautical Science, The Graduate University for Advanced Studies (SOKENDAI), Sagamihara City, Kanagawa 252-5210, Japan
[279] Andrews University, Berrien Springs, MI 49104, USA
[280] Research Center for Space Science, Advanced Research Laboratories, Tokyo City University, Setagaya, Tokyo 158-0082, Japan
[281] Institute for Cosmic Ray Research (ICRR), Research Center for Cosmic Neutrinos (RCCN), The University of Tokyo, Kashiwa City, Chiba 277-8582, Japan
[282] Department of Physics, Kyoto University, Sakyou-ku, Kyoto City, Kyoto 606-8502, Japan
[283] Yukawa Institute for Theoretical Physics (YITP), Kyoto University, Sakyou-ku, Kyoto City, Kyoto 606-8502, Japan
[284] Dipartimento di Scienze Aziendali—Management and Innovation Systems (DISA-MIS), Università di Salerno, I-84084 Fisciano, Salerno, Italy
[285] Van Swinderen Institute for Particle Physics and Gravity, University of Groningen, 9747 AG Groningen, The Netherlands
[286] Faculty of Science, Department of Physics, The Chinese University of Hong Kong, Shatin, N.T., Hong Kong
[287] Vrije Universiteit Brussel, B-1050 Brussel, Belgium
[288] Applied Research Laboratory, High Energy Accelerator Research Organization (KEK), Tsukuba City, Ibaraki 305-0801, Japan
[289] Department of Communications Engineering, National Defense Academy of Japan, Yokosuka City, Kanagawa 239-8686, Japan
[290] Department of Physics, University of Florida, Gainesville, FL 32611, USA
[291] Department of Information and Management Systems Engineering, Nagaoka University of Technology, Nagaoka City, Niigata 940-2188, Japan
[292] Tata Institute of Fundamental Research, Mumbai 400005, India
[293] Eindhoven University of Technology, 5600 MB Eindhoven, The Netherlands
[294] Department of Physics and Astronomy, Sejong University, Gwangjin-gu, Seoul 143-747, Republic of Korea

²⁹⁵ Concordia University Wisconsin, Mequon, WI 53097, USA
²⁹⁶ Department of Electrophysics, National Yang Ming Chiao Tung University, Hsinchu, Taiwan
²⁹⁷ Department of Physics, Rikkyo University, Toshima-ku, Tokyo 171-8501, Japan

## Abstract

We present the results of a model-based search for continuous gravitational waves from the low-mass X-ray binary Scorpius X-1 using LIGO detector data from the third observing run of Advanced LIGO and Advanced Virgo. This is a semicoherent search that uses details of the signal model to coherently combine data separated by less than a specified coherence time, which can be adjusted to balance sensitivity with computing cost. The search covered a range of gravitational-wave frequencies from 25 to 1600 Hz, as well as ranges in orbital speed, frequency, and phase determined from observational constraints. No significant detection candidates were found, and upper limits were set as a function of frequency. The most stringent limits, between 100 and 200 Hz, correspond to an amplitude $h_0$ of about $10^{-25}$ when marginalized isotropically over the unknown inclination angle of the neutron star's rotation axis, or less than $4 \times 10^{-26}$ assuming the optimal orientation. The sensitivity of this search is now probing amplitudes predicted by models of torque balance equilibrium. For the usual conservative model assuming accretion at the surface of the neutron star, our isotropically marginalized upper limits are close to the predicted amplitude from about 70 to 100 Hz; the limits assuming that the neutron star spin is aligned with the most likely orbital angular momentum are below the conservative torque balance predictions from 40 to 200 Hz. Assuming a broader range of accretion models, our direct limits on gravitational-wave amplitude delve into the relevant parameter space over a wide range of frequencies, to 500 Hz or more.

*Unified Astronomy Thesaurus concepts:* Gravitational waves (678); Gravitational wave astronomy (675); Low-mass x-ray binary stars (939); Neutron stars (1108)

## 1. Introduction

Rapidly rotating neutron stars (NSs) are primary targets for continuous gravitational-wave (GW) searches with the current network of ground-based detectors, LIGO, Virgo, and KAGRA. In these stars a deformation, or "mountain," sustained by elastic or magnetic strains, may result in a time-varying quadrupole from rotation, leading to the emission of GWs. Similarly, modes of oscillation may also lead to GW emission (see Lasky 2015 for a review).

In particular, NSs in low-mass X-ray binaries (LMXBs) are some of the most promising sources. In these systems magnetically channeled accretion from the companion onto the NS provides a mechanism to create a "mountain" (Ushomirsky et al. 2000; Melatos & Payne 2005; Osborne & Jones 2020; Singh et al. 2020), and the resulting GW torque may provide the solution to an astrophysical conundrum. There appears to be a sharp observed cutoff in the spin frequency ($\nu_s$) distribution of NSs in LMXBs at $\nu_s \approx 750$ Hz (Chakrabarty et al. 2003; Patruno et al. 2017), well below the theoretical breakup frequency for an NS (Haskell et al. 2018).

Although there are still several uncertainties in the modeling of the spin-up accretion torques (Patruno & Watts 2021; Glampedakis & Suvorov 2021), which may explain this observation (Patruno et al. 2012; Ertan & Alpar 2021), it has been suggested that the spin-down GW torques due to mountains can lead to an equilibrium at high frequencies that naturally explains the observed spins of NSs in LMXBs (Papaloizou & Pringle 1978; Wagoner 1984; Bildsten 1998) and the clustering of systems above $\nu_s \approx 500$ Hz and close to the maximum frequency (Patruno et al. 2017; Gittins & Andersson 2019). In such a scenario there is a natural correlation between the observed X-ray flux and the expected strength of the GWs, as a higher accretion rate leads to a stronger spin-up torque and thus requires a stronger GW torque for equilibrium (note that even if this equilibrium holds on average, there is expected to be some slight fluctuation in frequency or "spin wandering"; Bildsten et al. 1997; Mukherjee et al. 2018). Scorpius X-1 (Sco X-1), the most luminous LMXB, which is presumed to consist of an NS of mass $\approx 1.4 M_\odot$ in a binary orbit with a companion star of mass $\approx 0.4 M_\odot$ (Steeghs & Casares 2002), is therefore a very promising potential source of GWs. Some of the parameters inferred from electromagnetic observations of the system are summarized in Table 1. Note that the orbital eccentricity of Sco X-1 is believed to be small (Steeghs & Casares 2002; Wang et al. 2018) and is ignored in this search. Inclusion of eccentric orbits would add two search parameters that are determined by the eccentricity and the argument of periapse (Messenger 2011; Leaci & Prix 2015).

Given its promise as a source for potentially detectable continuous gravitational waves, Sco X-1 has been the subject of numerous GW searches and search methods to date, starting with a fully coherent search (Jaranowski et al. 1998) of 6 hr from initial LIGO's second science run (Abbott et al. 2007a). Beginning with the fourth science run, results for Sco X-1 have been reported (Abbott et al. 2007b; Abadie et al. 2011) as part of a search for stochastic signals from isolated sky positions, also known as the radiometer search (Ballmer 2006), which has continued in Advanced LIGO's first three observing runs (Abbott et al. 2017a, 2019, 2021a). Sco X-1 has also been included in a search principally designed for unknown binary systems (Goetz & Riles 2011) and subsequently improved to search directly for Sco X-1 (Meadors et al. 2016); these

---

²⁹⁸ lsc-spokesperson@ligo.org
²⁹⁹ Deceased, December 2021.
³⁰⁰ Deceased, November 2022.
³⁰¹ virgo-spokesperson@ego-gw.it
³⁰² Deceased, March 2022.
³⁰³ kscboard-chair@icrr.u-tokyo.ac.jp

**Table 1**
Observed Parameters of the LMXB Sco X-1

| Parameter | Value |
| --- | --- |
| Right ascension[a] | $16^h19^m55.0850^s$ |
| Decl.[a] | $-15^{[\circ]}38'24.9''$ |
| Distance (kpc) | $2.8 \pm 0.3$ |
| Orbital inclination $i$[b] | $44(\deg) \pm 6(\deg)$ |
| $K_1$ (km s$^{-1}$)[c] | [40, 90] |
| $t_{\rm asc}$ (GPS s)[d] | $974{,}416{,}624 \pm 50$ |
| $P_{\rm orb}$ (s)[e] | $68{,}023.86 \pm 0.043$ |

**Notes.** Uncertainties are $1\sigma$ unless otherwise stated. There are uncertainties (relevant to the present search) in the projected velocity amplitude $K_1$ of the NS, the orbital period $P_{\rm orb}$, and the time $t_{\rm asc}$ at which the NS crosses the ascending node (moving away from the observer), measured in the solar system barycenter.

[a] The sky position (as quoted in Abbott et al. 2007a, derived from Bradshaw et al. 1999) is determined to the microarcsecond and therefore can be treated as known in the present search.
[b] The inclination $i$ of the orbit to the line of sight, from observation of radio jets in Fomalont et al. (2001), is not necessarily the same as the inclination angle $\iota$ of the NS's spin axis, which determines the degree of polarization of the GW in Equation (1).
[c] The projected orbital velocity $K_1$ as estimated by Doppler tomography measurements and Monte Carlo simulations in Wang et al. (2018), which show $K_1$ to be weakly determined beyond the constraint that $40\ {\rm km\ s^{-1}} \lesssim K_1 \lesssim 90\ {\rm km\ s^{-1}}$.
[d] The time of ascension $t_{\rm asc}$, at which the NS crosses the ascending node (moving away from the observer), measured in the solar system barycenter, is derived from the time of inferior conjunction of the companion given in Wang et al. (2018) by subtracting $P_{\rm orb}/4$. It corresponds to a time of 2010 November 21 23:16:49 UTC and can be propagated to other epochs by adding an integer multiple of $P_{\rm orb}$, which results in increased uncertainty in $t_{\rm asc}$ and correlations between $P_{\rm orb}$ and $t_{\rm asc}$; see Figure 2.
[e] The orbital period reported in Wang et al. (2018). Note that this differs from the previous estimate in Galloway et al. (2014) by $2.6\sigma$.

**References.** Bradshaw et al. (1999); Fomalont et al. (2001); Wang et al. (2018).

searches were applied to data from initial LIGO's fifth and sixth science runs (Aasi et al. 2014; Meadors et al. 2017). A search method connected to Doppler-modulated sidebands (Messenger & Woan 2007; Sammut et al. 2014) was developed and applied to data from initial LIGO's sixth science run (Aasi et al. 2015a). This was further adopted into the so-called Viterbi search (Suvorova et al. 2016, 2017), which uses a hidden Markov model to track possible spin wandering; the Viterbi search has been applied to data from Advanced LIGO's first three observing runs (Abbott et al. 2017b, 2022a). The cross-correlation method (Dhurandhar et al. 2008; Whelan et al. 2015) used in the present work is an extension of the radiometer search that uses the signal model of GWs from an LMXB such as Sco X-1 to look for correlations between data at different times. It has been applied to data from Advanced LIGO's first and second science runs to set the strongest limits so far on GWs from Sco X-1 (Abbott et al. 2017c; Zhang et al. 2021).

The Advanced LIGO Gravitational-Wave Observatory (Aasi et al. 2015b) has conducted three observing runs, the last two in coordination with Advanced Virgo (Acernese et al. 2015). In these three runs, transient GWs were detected from over 90 coalescences of binary systems of black holes and/or NSs (Abbott et al. 2021b). The LIGO-Virgo O3 observing run (Buikema 2020; Abbott et al. 2020) began on 2019 April 01 15:00:00 UTC (GPS 1238166018), continued until a commissioning break at 2019 October 01 15:00:00 UTC (GPS 1253977218), resumed on 2019 November 01 15:00:00 UTC (GPS 1256655618), and ended on 2020 March 27 17:00:00 UTC (GPS 1269363618). In 2020 April, immediately following the LIGO-Virgo run, the KAGRA detector (Aso et al. 2013; Akutsu et al. 2021) and the GEO 600 detector (Lück et al. 2010; Affeldt et al. 2014; Dooley et al. 2016) conducted joint observations (Abbott et al. 2022b). In this analysis, we use data from the two LIGO detectors, as Virgo and KAGRA data were significantly less sensitive.

We use the calibrated data that are limited to times when a detector was in scientific observing mode (Davis et al. 2021). Due to the presence of transient instrumental glitches that degrade the sensitivity by raising the overall noise spectrum, we apply the "self-gating" procedure (Zweizig & Riles 2021) to remove these glitches when analyzing data below 600 Hz. This reduces the total volume of data included in the analysis below 600 Hz from 243–244 days to 231–240 days for the LIGO Hanford detector and from 250–251 days to 216–248 days for the LIGO Livingston detector. (The ranges are due to differences in the time baseline used in producing Fourier transforms at different frequencies; see Section 3.)

In addition, as in Abbott et al. (2017c), we exclude from our analysis frequencies at which the data are known to be influenced by instrumental disturbances of narrow frequency extent, known as "lines." In practice, this procedure removes data from times at which the signal model has Doppler-shifted the GW signal frequency $f_0$ of the search template into the instrumental line, reducing the sensitivity of the search near known lines.

The remainder of the paper is laid out as follows: In Section 2 we describe the properties of Sco X-1 and the modeled GW signal from it. In Section 3 we describe the specifics of the cross-correlation search as implemented for this analysis. In Section 4 we describe the identification and follow-up of potential signals. Section 5 sets upper limits on the strength of GWs from Sco X-1 from the sensitivity and result of the search on Advanced LIGO data and simulated signals and considers their implications on various torque balance models. Finally, Section 6 contains the conclusions.

## 2. Model of Gravitational Waves from Sco X-1

The modeled GW signal from a rotating NS consists of a "plus" polarization component $h_+(t) = A_+ \cos[\Phi(t)]$ and a "cross" polarization component $h_\times(t) = A_\times \sin[\Phi(t)]$.[304] The signal recorded in a particular detector will be a linear combination of $h_+$ and $h_\times$ determined by the detector's orientation as a function of time. The two polarization amplitudes are

$$A_+ = h_0 \frac{1 + \cos^2\iota}{2} \qquad \text{and} \qquad A_\times = h_0 \cos\iota, \qquad (1)$$

where $h_0$ is an intrinsic amplitude describing the strength of the signal when it reaches the solar system and $\iota$ is the inclination of the NS's spin to the line of sight. (For an NS in a binary, the spin inclination $\iota$ is not necessarily equal to the inclination $i$ of the binary orbit.) If $\iota = 0°$ or $180°$, $A_\times = \pm A_+$, and gravitational radiation is circularly polarized. If $\iota = 90°$, $A_\times = 0$, and it is linearly polarized. The general case, elliptical polarization, has $0 < |A_\times| < A_+$. Many search methods are sensitive to the combination

$$(h_0^{\text{eff}})^2 = \frac{A_+^2 + A_\times^2}{2} = h_0^2 \frac{[(1 + \cos^2 \iota)/2]^2 + [\cos \iota]^2}{2}, \quad (2)$$

which is equal to $h_0^2$ for circular polarization and $h_0^2/8$ for linear polarization (Messenger et al. 2015; this was the convention used in Abbott et al. 2017c but differs by a factor of 2.5 from the definition of $(h_0^{\text{eff}})^2$ in Whelan et al. 2015).

---

[304] The preferred (spin-2) polarization basis is constructed from orthonormal unit vectors in the plane of the sky: one along the projection of the NS spin axis, and the other along the intersection of the NS equatorial plane with the plane of the sky. This basis is rotated relative to a fiducial north-and-east-on-the-sky basis by an (unknown) polarization angle $\psi$, equivalent to the position angle of the NS's polarization axis (Jaranowski et al. 1998; Prix & Whelan 2007).

In order to understand the astrophysical relevance of the GW strengths we are probing, a useful benchmark is the so-called torque balance level. As already mentioned, it has been suggested that an LMXB, such as Sco X-1, may be in an equilibrium state where the spin-up torque due to accretion is balanced by a spin-down torque due to GW emission (Papaloizou & Pringle 1978; Wagoner 1984; Bildsten 1998). In order to obtain an estimate of the GW amplitude, we start by taking a simple spin-up torque of the form (Pringle & Rees 1972)

$$N_A = \dot{M}\sqrt{GMr}, \quad (3)$$

where $\dot{M}$ is the mass accretion rate onto the NS, which we infer from the X-ray flux $F_X$; $M$ is the mass of the star; $G$ is the gravitational constant; and $r$ is the lever arm, i.e., the radius at which the accretion torque is applied. By balancing the spin-up torque with the GW torque, i.e., imposing $N_A = \dot{E}_{GW}/2\pi\nu_s$, we can obtain the GW amplitude (Watts et al. 2008):

$$h_0 \approx 5.48 \times 10^{-27} \left(\frac{F_X}{10^{-8} \text{ erg cm}^{-2} \text{ s}^{-1}}\right)^{1/2} \left(\frac{\nu_s}{300 \text{ Hz}}\right)^{-1/2}$$
$$\times \left(\frac{R_*}{10 \text{ km}}\right)^{1/2} \left(\frac{r}{10 \text{ km}}\right)^{1/4} \left(\frac{M}{1.4 M_\odot}\right)^{1/4}. \quad (4)$$

Note that the usual torque balance benchmark assumes that accretion occurs at the surface of the NS, $r = R_* = 10$ km. (If the magnetic field is strong enough to truncate the accretion disk above the surface, the lever arm will instead be the Alfvén radius, $r = r_A$, and the GW amplitude implied by Equation (4) will be larger, as given in, e.g., Zhang et al. (2021) and Abbott et al. (2022a).) For Sco X-1, using the observed X-ray flux $F_X = 3.9 \times 10^{-7}$ erg cm$^{-2}$ s$^{-1}$ from Watts et al. (2008) and assuming that the GW frequency $f_0$ is twice the spin frequency $\nu_s$ (as would be the case for GWs generated by triaxiality in the NS), the torque balance value is

$$h_0 \approx 3.4 \times 10^{-26} \left(\frac{f_0}{600 \text{ Hz}}\right)^{-1/2}. \quad (5)$$

It is important to note that this amplitude is simply an order-of-magnitude estimate, which we use as a benchmark to understand whether our searches are probing astrophysically significant portions of parameter space. Much of the physics entering the accretion torque is, in fact, highly uncertain and depends on unknown physical parameters, such as the topology of the stellar magnetic field, the disk−field coupling, viscous heating in the disk, efficiency of X-ray emission, or radiation pressure in the disk. All these effects can strongly influence the spin-up torque, leading not only to a large rescaling (of up to an order of magnitude) of the strength of the torque in Equation (3) but in general also to different scalings with the parameters of the system (Patruno et al. 2012; Haskell et al. 2015; Glampedakis & Suvorov 2021). For example, Andersson et al. (2005) have even suggested that, for high accretion luminosities, radiation pressure will lead to a sub-Keplerian disk and a strongly reduced spin-up torque. In this case Sco X-1 would host a slowly rotating NS, which does not emit GWs in our current search band. In light of the various uncertainties, we will retain the standard simplifying assumptions in the derivation of Equation (5) for most of our torque balance comparisons, but keep in mind that the torque balance level is uncertain and model dependent and should not be interpreted too strictly.

### 3. Setup of Cross-correlation Search

The cross-correlation (CrossCorr) search method (Dhurandhar et al. 2008; Whelan et al. 2015) has been used to search for GWs from Sco X-1 in LIGO data from the first two observing runs of Advanced LIGO and Advanced Virgo (Abbott et al. 2017c; Zhang et al. 2021). It uses the signal model described in Section 2 to construct an appropriately weighted statistic $\rho$ including correlations between data separated by up to a coherence time $T_{\max}$. The statistic is constructed using short Fourier transforms (SFTs) of length $T_{\text{sft}}$. If the SFTs are labeled by an index $K$, $L$, etc., which encodes the detector and time of the SFT, and $z_K$ is an appropriately normalized combination of the Fourier data at the frequency of interest, we can write the statistic $\rho$ as

$$\rho = \sum_{KL \in \mathcal{P}} (W_{KL} z_K^* z_L + W_{KL}^* z_K z_L^*), \quad (6)$$

where $\mathcal{P}$ is the set of all pairs of SFTs whose start times differ by $T_{\max}$ or less and $W_{KL}$ is a complex weighting factor constructed using the signal model. Since the choice of frequency bin(s) in the construction of $z_K$ and the amplitude and phase of the weighting factor $W_{KL}$ for each SFT pair depend on the unknown parameters of the signal, we must conduct the search at a set of points in parameter space, each of which defines a "search template."

The maximum separation $T_{\max}$ can be chosen to "tune" the search: higher $T_{\max}$ values produce a more sensitive search but can significantly increase computing cost, both due to the increased number of correlation terms in the statistic and especially due to the increased density of search templates needed in the parameter space. As detailed in Abbott et al. (2017c), the $T_{\text{sft}}$ and $T_{\max}$ values were chosen as a function of GW frequency and orbital parameters in order to optimize the search at a given computing cost. For the present search, we used the same $T_{\max}$ values as in O1 rather than re-optimizing.[305] The one exception is for the GW frequency range 400–600 Hz, for which the achievable sensitivity is closer in O3 than it was in O1 to the signal strength nominally expected from the torque balance model Equation (5); for those frequencies, we used double the $T_{\max}$ of the O1 search. Note that, even with the same coherence times, the O3 search would require more computing

---

[305] The O2 analysis of Zhang et al. (2021), which was limited to GW frequencies between 40 and 180 Hz, used a longer coherence time of ∼19 hr by leveraging the resampling techniques described in Meadors et al. (2017).

**Table 2**
Parameters Used for the Cross-correlation Search

| Parameter | Range |
|---|---|
| $f_0$ (Hz) | [25, 1600] |
| $a \sin i$ (lt-s)[a] | [1.44, 3.25] |
| $t'_{asc}$ (GPS s)[b] | $1{,}255{,}015{,}049 \pm 3 \times 185$ |
| $\tilde{P}$ (s)[c] | $68{,}023.86 \pm 3.3 \times 0.011$ |

**Notes.**
[a] The range for the projected semimajor axis $a \sin i = K_1 P_{orb}/(2\pi)$ in lt-s was taken from the constraint $K_1 \in [40, 90]$ km s$^{-1}$.
[b] This value for the time of ascension $t'_{asc}$, defined in Equation (7), has been propagated forward by 4125 orbits from the value of $t_{asc}$ in Table 1 and corresponds to a time of 2019 October 13 15:17:11 UTC, near the middle of the O3 run. The increase in uncertainty is due to the uncertainty in $P_{orb}$.
[c] This is the "sheared" period defined in Equation (8); note that the uncertainty in $\tilde{P}$ has been reduced compared to the marginal uncertainty in $P_{orb}$ by the same factor by which the uncertainty in $t'_{asc}$ has been increased relative to that for $t_{asc}$, as described in Wagner et al. (2022). The search region in $\tilde{P}$ is given by the elliptical boundary shown in Figure 1, but it is defined as "unresolved" if only one template is needed to cover $68{,}023.86 \pm 3.3 \times 0.011$ at a maximum mismatch of 0.0625.

resources than the O1 search, due to the increased observing time. However, by using a more efficient template lattice and convenient coordinate choices as described in Wagner et al. (2022), we are able to offset the increase in observing time and maintain a manageable computing time.

In addition to the GW signal frequency, $f_0$, we search over the orbital parameters of the system, as summarized in Table 2. The projected semimajor axis of the orbit is assumed to lie in the range $a \sin i \in [1.44, 3.25]$ lt-s, corresponding to a range in projected orbital velocity of [40, 90] km s$^{-1}$. The search region in orbital period $P_{orb}$ and time of ascension $t_{asc}$ is constructed using the method of Wagner et al. (2022): the time of ascension $t_{asc}$ is propagated 4125 orbits to define

$$t'_{asc} = t_{asc} + 4125 \times 68023.86 \text{ s}, \quad (7)$$

and the "sheared" orbital period is defined as

$$\tilde{P} = P_{orb} - 2.42 \times 10^{-4}(t'_{asc} - 1255015049). \quad (8)$$

The most likely $t'_{asc}$ is 2019 October 13 15:17:11 UTC (GPS 1255015049). The coordinates $t'_{asc}$ and $\tilde{P}$ are approximately uncorrelated both in the parameter space metric of the search and in the astrophysical prior distribution.[306] Note that the included prior probability is an underestimate of the efficiency in covering parameter space, since the "sheared" period coordinate $\tilde{P}$ was unresolved in many search jobs, i.e., the mismatch associated with an offset of $3.3 \times 0.011$ s from the most likely value 68023.86 s (the dashed rectangular boundaries in Figure 1) was less than 0.0625.

The search was done over a range of $t'_{asc} = 1255015049 \pm 3 \times 185$ s and with $\tilde{P}$ constrained to lie in an elliptical region centered on 68,023.86 s with semiaxes of $3.3 \times 0.011$ s for $\tilde{P}$ and $3.3 \times 185$ s for $t'_{asc}$, as illustrated in Figure 1. For reference, in Figure 2 we show this region in the coordinates $(t'_{asc}, P_{orb})$, along with the search regions used in Abbott et al. (2017c) and Zhang et al. (2021), propagated forward in time to the epoch considered in this analysis.

To perform the analysis, the parameter space was divided into small jobs that could be run in parallel. Each 5 Hz GW frequency band was divided into between 100 and 4000 sub-bands, and the orbital parameter space was divided into 9 or 20 cells, as illustrated in Figure 3. These subdivisions of parameter space were chosen so that each analysis job would run in a reasonable amount of time ($\lesssim$10 hr), allowing the analysis to be done quickly via distributed computing.

Search templates were placed in the parameter space using an $A_n^*$ lattice with a maximum mismatch of 0.25 as described in Wagner et al. (2022) and Wette (2014). Because the use of the sheared coordinate $\tilde{P}$ reduces the 1$\sigma$ prior uncertainty from 0.043 to 0.011 s, the period was unresolved for most search jobs. We computed the maximum mismatch $\mu_{max}^{\tilde{P}}$ associated with an offset of $3.3 \times 0.011$ s. For jobs in which $\mu_{max}^{\tilde{P}} \leqslant 0.0625$, the initial search was done with $\tilde{P} = 68023.86$ s and an $A_3^*$ lattice in the other three parameters $(f_0, a \sin i, t'_{asc})$. To account for the nonzero mismatch contribution from the possible $\tilde{P}$ offset, the maximum mismatch of the $A_3^*$ lattice was set to $0.25 - \mu_{max}^{\tilde{P}}$. (Note that this actually guarantees coverage at the specified maximum mismatch over a rectangle in $t'_{asc}, \tilde{P}$ rather than just the ellipse shown in Figure 1.) If, on the other hand, the $\tilde{P}$ coordinate was resolved in a particular search job, an $A_4^*$ lattice in all of the coordinates with maximum mismatch of 0.25 was used.

As noted in Section 1, even in approximate equilibrium, Sco X-1 may undergo stochastic variation of the GW frequency $f_0$, also known as "spin wandering." As in Abbott et al. (2017c), we can apply the estimates in Whelan et al. (2015) to set bounds on the loss of signal-to-noise ratio (S/N) under a simplistic model in which the GW frequency undergoes a net spin-up or spin-down of magnitude $|\dot{f}|_{drift}$, changing on a timescale $T_{drift}$. For O3, in which the duration of the run from start to end is $T_{run} = 3.12 \times 10^7$ and the coherence times used for the initial search are $T_{max} \leqslant 18720$ s, the expected loss of S/N is

$$\frac{E[\rho]^{ideal} - E[\rho]}{E[\rho]^{ideal}}$$
$$\approx 0.018 \left(\frac{T_{drift}}{10^6 \text{ s}}\right)\left(\frac{|\dot{f}|_{drift}}{10^{-12} \text{ Hz s}^{-1}}\right)^2 \left(\frac{T_{max}}{18720 \text{ s}}\right)^2, \quad (9)$$

which indicates that spin wandering is not likely to be an important effect for this search. In addition, as noted in Zhang et al. (2021), the predictions of Mukherjee et al. (2018) based on the time variation of the X-ray flux from Sco X-1 imply considerably less spin wandering than the naive $|\dot{f}|_{drift} - T_{drift}$ model, whose fiducial parameters are taken from Messenger et al. (2015).

### 4. Candidates, Outliers, and Follow-up

The detection statistic $\rho$ is normalized to have zero mean and unit variance in Gaussian noise. As in Abbott et al. (2017c), we make a naive estimate of the expected background by assuming that each search template represents an independent Gaussian random number, and we use this value to set the threshold at an approximate level of one expected false-alarm level per 50 Hz. As shown in Figure 4, the threshold for follow-up for this search was set at 6.3 for 25 Hz $< f_0 <$ 400 Hz, 6.2 for

---

[306] The correlations would have been even smaller had the optimal coefficient $2.25 \times 10^{-4}$ been used in Equation (8), but a software bug led to the use of the coefficient $2.42 \times 10^{-4}$. The impact of this error is negligible, however, reducing the prior probability covered by the search region from 99.4% to 99.2%.

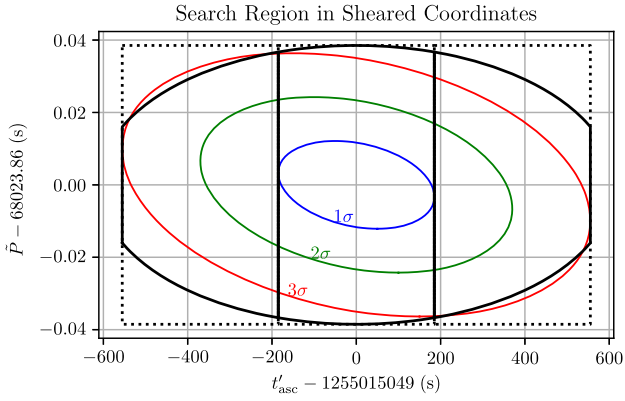

**Figure 1.** Search region in terms of parameters $t'_{\rm asc}$ and $\tilde{P}$ defined in Equations (7) and (8), respectively. The lattice is constructed to completely cover, with maximum mismatch 0.25, the solid black truncated ellipse. The solid black ellipse has semiaxes $3.3 \times 185$ s in $t'_{\rm asc}$ and $3.3 \times 0.011$ s in $\tilde{P}$, and the truncating boundaries are at $\pm 3 \times 185$ s of the most likely $t'_{\rm asc}$ value. For comparison, the thin colored ellipses show curves of constant prior probability corresponding to $1\sigma$, $2\sigma$, and $3\sigma$ (containing 39.3%, 86.5%, and 98.9% of the prior probability, respectively), including the effects of changing coordinates from $t_{\rm asc}$ and $P_{\rm orb}$ appearing in Table 1 to $t'_{\rm asc}$ and $\tilde{P}$. The inner search region, in which we choose a longer $T_{\rm max}$ to do a deeper search, is within $\pm 185$ s of the most likely value of $t'_{\rm asc}$ and contains 68.1% of the prior probability, while the full search region, within $\pm 3 \times 185$ s of the most likely $t'_{\rm asc}$ value, contains 99.2% of the prior probability. The slight misalignment of the prior and search ellipses is due to a software bug, which led to a definition of $\tilde{P}$ that differed slightly from the optimal one, as described in Section 3. The dashed rectangular boundaries show the region effectively covered by the majority of search jobs for which the "sheared" period coordinate $\tilde{P}$ was unresolved in many search jobs, i.e., the mismatch associated with an offset of $3.3 \times 0.011$ s from the most likely value 68023.86 s was less than 0.0625.

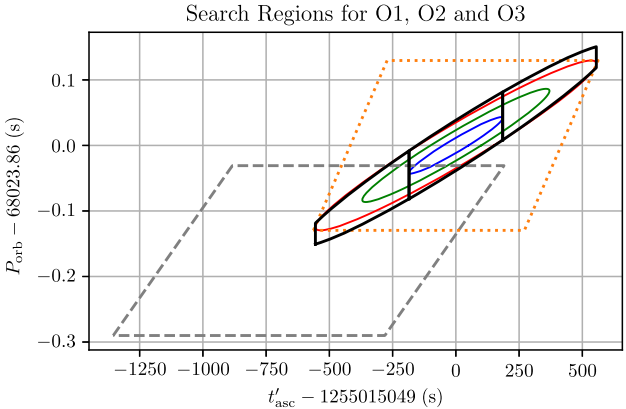

**Figure 2.** The search regions and prior uncertainties shown in Figure 1, expressed in terms of the system parameters $t'_{\rm asc}$ and $P_{\rm orb}$. For reference, we also show the regions used for the O1 analysis in Abbott et al. (2017c; gray dashed lines) and reported for the O2 analysis in Zhang et al. (2021; orange dotted lines), propagated to the epoch of this search (which transforms the rectangular search regions into parallelograms). Note that the search region for Abbott et al. (2017c) is offset in both $P_{\rm orb}$ and $t'_{\rm asc}$ because it used the $P_{\rm orb}$ estimate in Galloway et al. (2014), while the others used the updated estimate in Wang et al. (2018). The analysis of Abbott et al. (2017c) is still believed to have covered the plausible signal space because of the underresolution of the period direction and the fact that the offset in $t'_{\rm asc}$ induced by the inaccurate $P_{\rm orb}$ value was less for the epoch of the search (2015 rather than 2019).

400 Hz $< f_0 <$ 600 Hz, and 5.8 for 600 Hz $< f_0 <$ 1600 Hz. Due in part to our more efficient template placement, we were able to use a lower follow-up threshold than in Abbott et al. (2017c), except for 400 Hz $< f_0 <$ 600 Hz, where we used the same threshold, despite having a search with twice the coherence

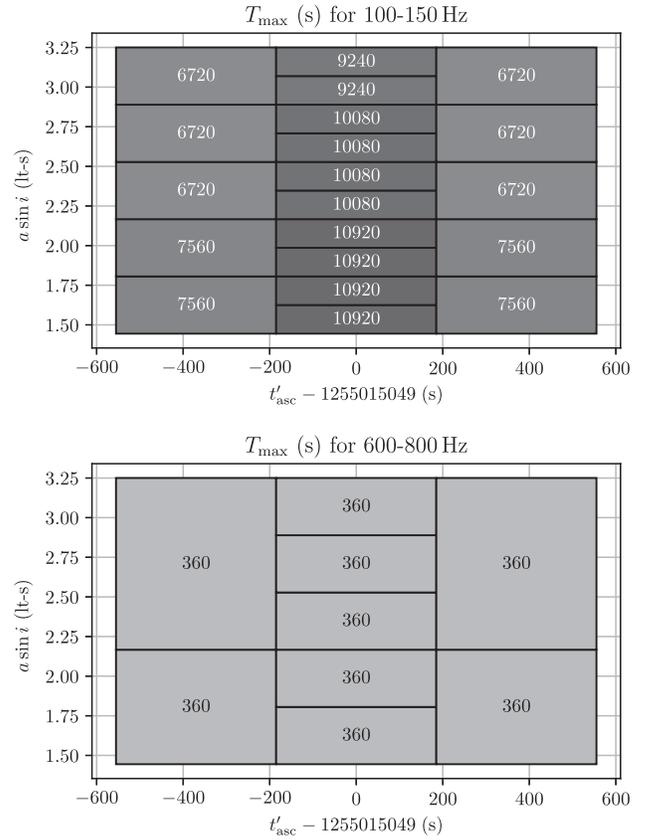

**Figure 3.** Illustration of parameter space cells in $t'_{\rm asc}$ and $a \sin i$ and example of coherence times $T_{\rm max}$, in seconds, chosen as a function of the orbital parameters of the NS. Increasing coherence time improves the sensitivity but increases the computational cost of the search. The values used are the same as in Abbott et al. (2017c), except between 400 and 600 Hz, where they have been doubled; see Table 3 for details.

time. Table 3 shows the resulting expected numbers of false alarms in each GW frequency range.

For candidates exceeding the follow-up threshold, we applied the procedure detailed in Abbott et al. (2017c). Here we highlight the basic steps, as well as details that were changed for this analysis:

1. Candidates were "clustered" together in GW frequency, with all templates within 0.01 Hz of a peak in S/N above the threshold being represented by the parameters of the peak. These are known as the "level 0" results.

2. A "refinement" search was performed on each level 0 candidate, with the same $T_{\rm max}$ as the original search and a resolution $\sim 3\times$ as fine as the original lattice. This and later stages of follow-up were run on a rectangular grid in the "sheared" parameters $(f_0, a \sin i, t'_{\rm asc}, \tilde{P})$. For the refinement stage, a grid of $13 \times 13 \times 13 \times 5$ points was used, centered on the $(f_0, a \sin i, t'_{\rm asc})$ values of the candidate, and covering the full prior range in the initially unresolved $\tilde{P}$. To deal with the effects of unknown narrowband features ("lines") present in a single detector, we computed a detection statistic using only data from the LIGO Livingston Observatory (LLO) detector and another using only LIGO Hanford Observatory (LHO) data. If either of these exceeded the detection statistic constructed from all the data, we vetoed the candidate as a likely instrumental artifact. Candidates from the

Table 3
Summary of Numbers of Templates and Candidates

| $f_0$ (Hz) Min | $f_0$ (Hz) Max | $T_{sft}$ (s) | $T_{max}$ (s) Min | $T_{max}$ (s) Max | $\rho$ Thresh[a] | Number of Templates | Expected Gauss False Alarms[b] | Follow-up Level 0[c] | 1[d] | 2[e] | 3[f] |
|---|---|---|---|---|---|---|---|---|---|---|---|
| 25 | 50 | 1440 | 10080 | 18720 | 6.3 | $5.68 \times 10^9$ | 0.8 | 63 | 31 | 10 | 1 |
| 50 | 100 | 1020 | 8160 | 14280 | 6.3 | $2.88 \times 10^{10}$ | 4.3 | 131 | 114 | 40 | 2 |
| 100 | 150 | 840 | 6720 | 10920 | 6.3 | $6.13 \times 10^{10}$ | 9.1 | 169 | 166 | 73 | 3 |
| 150 | 200 | 720 | 5040 | 8640 | 6.3 | $6.69 \times 10^{10}$ | 10.0 | 171 | 170 | 68 | 2 |
| 200 | 300 | 600 | 2400 | 4800 | 6.3 | $4.54 \times 10^{10}$ | 6.7 | 66 | 66 | 14 | 4 |
| 300 | 400 | 510 | 1530 | 3060 | 6.3 | $2.09 \times 10^{10}$ | 3.1 | 19 | 19 | 1 | 1 |
| 400 | 600 | 360 | 720 | 2160 | 6.2 | $3.46 \times 10^{10}$ | 9.8 | 343 | 226 | 20 | 9 |
| 600 | 800 | 360 | 360 | 360 | 5.8 | $1.80 \times 10^9$ | 6.0 | 15 | 15 | 0 | 0 |
| 800 | 1200 | 300 | 300 | 300 | 5.8 | $4.36 \times 10^9$ | 14.5 | 226 | 70 | 2 | 0 |
| 1200 | 1600 | 240 | 240 | 240 | 5.8 | $4.36 \times 10^9$ | 14.5 | 346 | 55 | 2 | 0 |

**Notes.** For each range of GW frequencies, this table shows the SFT duration $T_{sft}$; the minimum and maximum coherence time $T_{max}$ used for the search, across the different orbital parameter space cells (see Figure 3); the threshold in S/N $\rho$ used for follow-up; the total number of templates; and the number of candidates at various stages of the process. (See Section 4 for detailed description of the follow-up procedure.)
[a] This is the threshold for initiating follow-up, i.e., to produce a level 0 candidate.
[b] This is the number of candidates that would be expected in Gaussian noise, given the number of templates and the follow-up threshold.
[c] This is the actual number of candidates (after clustering) that crossed the S/N threshold and were followed up.
[d] This is the number of candidates remaining after refinement. All of the candidates "missing" at this stage have been removed by the single-detector veto for unknown lines, defined in Section 4.
[e] This is the number of candidates remaining after each has been followed up with a $T_{max}$ equal to 4× the original $T_{max}$ for that candidate. (True signals should approximately double their S/N; any candidates whose S/N goes down have been dropped.) All of the signals present at this stage are shown in Figure 5, which also shows the behavior of the search on simulated signals injected in software.
[f] This is the number of candidates remaining after $T_{max}$ has been increased to 16× its original value.

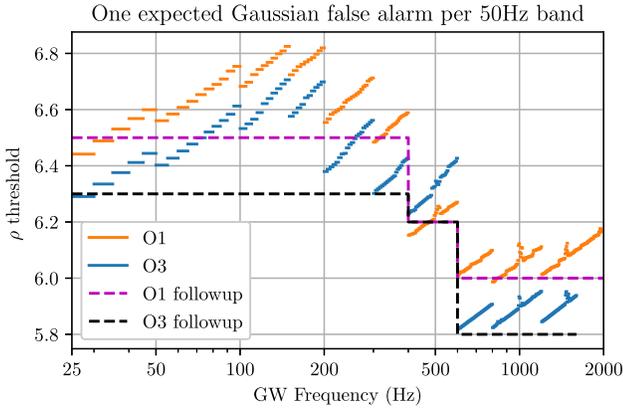

**Figure 4.** Selection of follow-up threshold as a function of GW frequency. If the data contained no signal and only Gaussian noise, each template in the parameter space would have some chance of producing a statistic value exceeding a given threshold. Within each 5 Hz frequency band, the total number of templates was computed and used to find the threshold at which the expected number of Gaussian outliers above that value would be 0.1. This is shown with short blue lines for the templates in the present search (see Table 3); for reference, the thresholds calculated from the numbers of templates in the O1 search of Abbott et al. (2017c) are shown in orange. Because of the more efficient template placement of algorithm from Wagner et al. (2022), the O3 search has fewer templates, and therefore a lower implied threshold, than the O1 search, which used the same coherence times. The exception is for 400 Hz < $f_0$ < 600 Hz, where the same threshold of 6.2 was used for the O1 and O3 searches, and the latter used twice the coherence time (and therefore a denser parameter space lattice) as the former. The present search uses a threshold of 6.3 for 25 Hz < $f_0$ < 400 Hz and 5.8 for 600 Hz < $f_0$ < 1600 Hz (black dashed line), which are lower than in the O1 search (magenta dashed line). Note that the large number of non-Gaussian outliers (see Table 3) makes the Gaussian follow-up level an imprecise tool in any event.

refinement search that survive this veto are known as the "level 1" results.

3. Two successive rounds of follow-up were performed, starting with the level 1 candidates. At each stage, the coherence time $T_{max}$ was quadrupled from the previous stage, and the density of templates in each direction was increased by a factor of 3. The grid used was $13 \times 13 \times 13 \times 13$ in ($f_0$, $a \sin i$, $t'_{asc}$, $P$), centered on the peak of the previous level's results. Candidates that increased their S/N from the previous level were known as the "level 2" (4× the original $T_{max}$) and "level 3" (16× the original $T_{max}$) results.

A total of 22 candidates survive level 3 of follow-up. Two checks were done to determine whether any of them represent convincing detection candidates: one using the results of the cross-correlation search, and one using an independent pipeline.

For the first check, in Figure 5 we plot the ratio by which the S/N increases from level 1 to level 2 and from level 2 to level 3. We also plot the corresponding ratios for all of the candidates surviving level 2 (the 16× original $T_{max}$ follow-up is not available for candidates that fail level 2), as well as for the simulated signal injections described in Section 5. We would naively expect a real signal to double its S/N between level 1 and level 2, and again between level 2 and level 3, but none of the candidates from the search come close to this. As in Abbott et al. (2017c), none of the candidates double their S/N from level 1 to level 3, let alone in a single follow-up stage. On the other hand, all but one of the injections, while not doubling their S/N with each stage of follow-up, increase their S/N noticeably more than any of the candidates from the search.

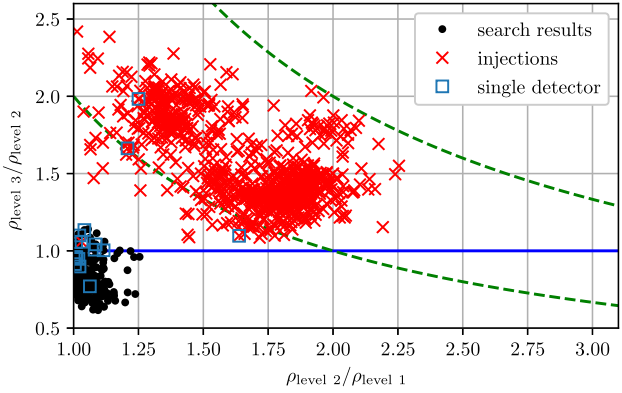

**Figure 5.** Ratios of follow-up statistics for search candidates and simulated signals. This plot shows all of the candidates that survived level 2 of follow-up (see Section 4 and Table 3), both from the main search and from the analysis of the simulated signal injections described in Section 5. It shows the ratios of the S/N $\rho$ after follow-up level 1 (at the original coherence time $T_{\max}$), level 2 (at $4\times$ the original coherence time), and level 3 (at $16\times$ the original coherence time). (The boxes labeled "single detector" are outliers or injections at GW frequencies where only one detector's data were included in the analysis because of known instrumental artifacts in the other detector.) The green dashed lines are at constant values of $\rho_{\text{level 3}}/\rho_{\text{level 1}}$ equal to 2 and 4. There are no points with $\rho_{\text{level 2}}/\rho_{\text{level 1}} < 1$ because those candidates do not survive level 2 follow-up and are therefore not subjected to level 3 follow-up. From the construction of the statistic in Whelan et al. (2015), the naive expectation is that the S/N will roughly double each time $T_{\max}$ is quadrupled. Empirically, the follow-ups of injections do not show exactly that relationship, but all but one (which was injected at a frequency contaminated by instrumental artifacts) show significant increases in S/N that are not seen in any of the follow-ups of search candidates. We thus conclude that no convincing detection candidates are present.

Note that, at some GW signal frequencies, known instrumental lines lead us to omit all the data from one of the two LIGO detectors from the search. This produces a single-detector search for which the unknown-line veto is not applicable. Of the 22 search outliers that survived our veto process, 6 were in this category, as well as 4 of the injections, including the one injection that increased its S/N negligibly under the follow-up procedure. Also note that of the 821 injected signals (out of 918) that produced $\rho$ values above their respective thresholds, 817 survived all the levels of follow-up. (There were three vetoed at level 1, one at level 2, and zero at level 3, all because of the single-detector unknown-line veto.) We thus conclude that our follow-up procedure is relatively robust and that there are no convincing detection candidates from the search.

An additional, complementary, multistage MCMC follow-up using the method described in Tenorio et al. (2021) using the PyFstat package (Ashton & Prix 2018; Keitel et al. 2021; Ashton et al. 2022) was applied to the 22 outliers using the same configuration as in Abbott et al. (2022c). This method places templates adaptively to compute the semicoherent $\mathcal{F}$-statistic (Jaranowski et al. 1998; Cutler & Schutz 2005) around the candidate of interest using a diminishing number of coherent segments (660, 330, 92, 24, 4, and 1). The coherence times of the corresponding segments range from half a day to the full observing run. A Bayes factor is computed using the $\mathcal{F}$-statistic values from subsequent coherence stages corresponding to the loudest template. The signal hypothesis assesses the consistency of these values, whereas the noise hypothesis states the (in)consistency of the final value with the background distribution, taking the final-stage factor into account.

The resulting Bayes factor values are significantly lower than what would be expected for a signal detectable by this search, as confirmed by an analogous follow-up of a similar number of injected signals.

## 5. Upper Limits and Implications

Since our search produced no convincing detection candidates, we set upper limits on the strength of GWs from Sco X-1 as a function of GW frequency, using the method described in detail in Abbott et al. (2017c). First, naive Bayesian upper limits were set within each 0.05 Hz band, using the 95th percentile of the posterior on $h_0^2$ or $(h_0^{\text{eff}})^2$ deduced from the highest S/N seen in each band.[307] Then, a series of simulated signals were added to the data at a variety of amplitudes, and a Bayesian logistic regression analysis was performed to estimate the factor by which to multiply the naive 95% Bayesian upper limit on amplitude in order to reach the threshold of 95% signal recovery. (As in Abbott et al. (2017c), "recovery" was defined as an increase in the maximum S/N seen in a band, over the value with no injection present.) We performed a total of 918 injections[308] between 25 and 500 Hz, of which 863 were recovered, with a resulting adjustment factor of 1.19 to the amplitude of the $h_0^{\text{eff}}$ upper limit. As described in Abbott et al. (2017c), to set the adjustment factor for the $h_0$ limit, we limit attention to injections that were generated with a specified $h_0$ rather than $h_0^{\text{eff}}$, of which we recovered 546 of 575, with a resulting adjustment factor of 1.17.[309] The upper limits including these adjustment factors are shown in Figure 6.

The upper limits placed by this search improve on those from previous observing runs and are now probing a theoretically significant portion of parameter space. This is usually quantified in terms of the torque balance amplitude, i.e., the GW amplitude that would be required for GW spin-down torques to balance the accretion torque. As discussed in Section 1, this limit is a useful benchmark but is highly uncertain, reflecting the high level of uncertainty in the theoretical modeling of accretion torques. We illustrate this in Figure 7 by comparing the upper limits from our search not only to the standard limit of Equation (5), obtained by assuming the lever arm to be the radius of the NS, $r = R_* = 10$ km, in Equation (4), but also to an example of the range of torque balance amplitudes allowed by current models for accretion onto a magnetized star. First, we consider a model of the same form as in Equations (3) and (5), but we do not assume accretion to occur on the surface, but rather take the torque arm to be the Alfvén radius $r_A$, at which the disk is truncated by the magnetic field (Pringle & Rees 1972):

$$r_A = 35\, X \left(\frac{\dot{M}}{10^{-10} M_\odot\, \text{yr}^{-1}}\right)^{-2/7} \left(\frac{M}{1.4 M_\odot}\right)^{-1/7}$$
$$\times \left(\frac{R}{10\,\text{km}}\right)^{12/7} \left(\frac{B}{10^8\,\text{G}}\right)^{4/7}\,\text{km}, \qquad (10)$$

---

[307] We used a simple extreme value likelihood assuming independent Gaussian distributions for the detection statistics from the templates in the initial bank. Future work may leverage more sophisticated methods of estimating this distribution, such as those of Tenorio et al. (2022).
[308] These are the same injections that were used to validate the follow-up procedure, as described in Section 4.
[309] For comparison, the adjustment factors in Abbott et al. (2017c) were 1.44 for $h_0$ and 1.21 for $h_0^{\text{eff}}$. The fact that the factors derived from the current search are comparable is evidence that the template bank modifications of Wagner et al. (2022) do not significantly reduce the sensitivity of the search.

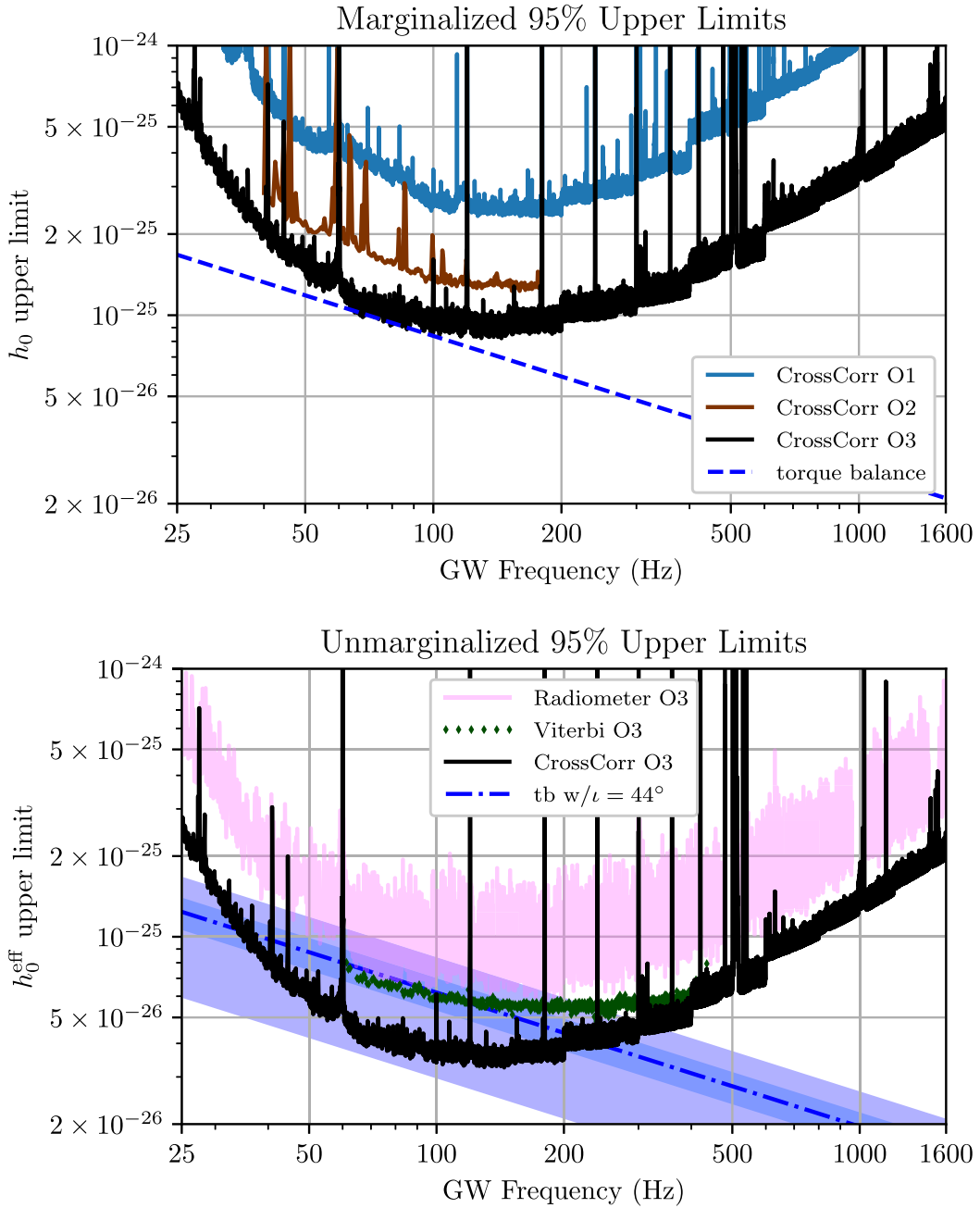

**Figure 6.** Upper limits from directed searches in advanced LIGO data. Top: upper limit on $h_0$, after marginalizing over NS spin inclination $\iota$, assuming an isotropic prior. The dashed line shows the nominal expected level assuming torque balance (Equation (5)) as a function of GW frequency. Bottom: upper limit on $h_0^{\rm eff}$, defined in Equation (2). This is equivalent to the upper limit on $h_0$ assuming circular polarization. (Note that the marginalized upper limit in the top panel is dominated by linear polarization and so is a factor of almost $\sqrt{8}$ higher). The blue dotted–dashed line (labeled as "tb w/$\iota = 44°$") corresponds to the assumption that the NS spin is aligned to the most likely orbital angular momentum and $\iota \approx i \approx 44°$ (see Table 1). The blue diagonal bands show $h_0^{\rm eff}$ levels corresponding to the torque balance $h_0$ in the top panel. The darker-shaded band corresponds (5th to 95th percentiles) to a Gaussian distribution with mean and standard deviation corresponding to $\iota = 44° \pm 6°$, as used in Zhang et al. (2021). Finally, the lighter-shaded band shows the full range of possible $h_0^{\rm eff}$ values corresponding to torque balance, with circular polarization at the top and linear polarization on the bottom. For comparison with the "CrossCorr O3" results presented in this paper, we show in the top panel the isotropic marginalized limits from the previous cross-correlation searches in Abbott et al. (2017c) ("CrossCorr O1") and Zhang et al. (2021) ("CrossCorr O2"). In the bottom panel we include the limits assuming circular polarization from other searches of O3 data: "Radiometer O3" is the narrowband radiometer analysis of Abbott et al. (2021a), which used data from Advanced LIGO's first three observing runs, and "Viterbi O3" is the analysis of Abbott et al. (2022a) using a hidden Markov model.

where $0.1 \lesssim X \lesssim 1$ is a phenomenological parameter that encodes the uncertainty in the truncation radius of the disk. Using the mass accretion rate of Sco X-1 inferred from X-ray observations (Watts et al. 2008), along with $B = 10^8$ G and $X = 1$, gives an Alfvén radius of $r_A \approx 49$ km, which is used to generate the curve in Figure 7. We also consider one of the parameterized models of Glampedakis & Suvorov (2021), which encompass a wide range of physics and can successfully fit spin-up episodes in a number of observed LMXBs. In particular, for illustrative purposes, we use their "new" model 2, for which the torque has the form $N_A^{(2)} = \xi^{-7/2} N_A (1 + 3\xi^{7/2} - 2\omega_A)/3$, with the fastness parameter $\omega_A = (R_A/R_C)^{(3/2)}$ and the corotation radius $R_c = 27(M/1.4M_\odot)^{(1/3)}(\nu_s/500)^{(-2/3)}$ km. We consider two values for the magnetic field strength at the surface of the star, $B = 10^8$ and $10^9$ G, and consider values of $\xi$ between 0.3 (which sets the upper limit in our plots) and 0.5.

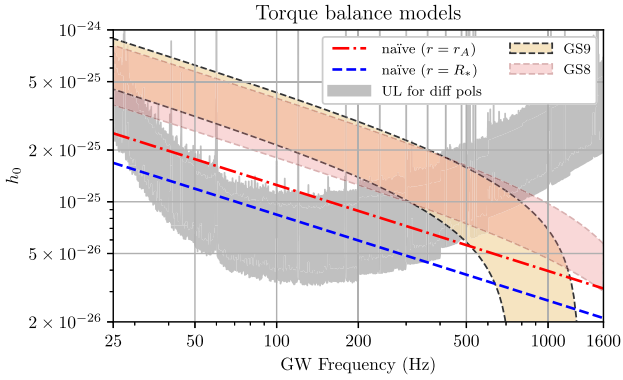

**Figure 7.** Comparison of upper limits to predictions of torque balance models. The gray band indicates the $h_0$ upper limit implied by the $h_0^{\text{eff}}$ upper limit in the bottom panel of Figure 6, assuming the range of possible inclinations from $0 \leqslant |\cos \iota| \leqslant 1$. (Linear polarization ($\cos \iota = 0$) is at the top, and circular polarization ($\cos \iota = \pm 1$) is at the bottom.) The dashed blue line is the usual conservative torque balance estimate assuming accretion at the surface of the NS, $r = R_* = 10$ km. The dotted–dashed red line is the same model assuming a lever arm of $r = r_A \approx 49$ km, which is the value of the Alfvén radius given by Equation (10) for $\xi = 1$ and $B = 10^8$ G. Note that this is slightly larger than the value of 35 km used in, e.g., Zhang et al. (2021) and Abbott et al. (2022a), since we use the inferred mass accretion rate of Sco X-1. The colored bands show the range of predictions for the models of Glampedakis & Suvorov (2021), in particular model 2, for the range of $0.3 < \xi < 0.5$, assuming a magnetic field of $10^8$ G (GS8) or $10^9$ G (GS9).

As can be clearly seen, there is a wide portion of parameter space allowed by theoretical models. Furthermore, at higher frequencies the theory is intrinsically uncertain; it is in fact unclear whether one should expect the source to be rapidly rotating, as some of the models of Glampedakis & Suvorov (2021) predict spin equilibrium due to accretion alone below GW frequencies of roughly 1 kHz, without any need for gravitational radiation—in fact, it is also possible that the NS may not be in the frequency range we are searching. This "fuzziness" in the torque balance limit (which may be even further enhanced by additional effects not considered in the models of Glampedakis & Suvorov (2021), such as viscous heating in the disk or the efficiency of X-ray emission) makes it therefore impossible to draw firm conclusions on the equation of state (EOS) of the NS, or magnetic field strength, directly from our upper limits, without committing to a particular accretion model. It is therefore important to remember, when analyzing our results, that torque balance may not be active in this system, or that even if it is, the GW torques may only be active for a fraction of the time, with a low duty cycle for GW emission (Haskell et al. 2015).

Nevertheless, it is clear from Figure 7 that in the range we are most sensitive to, approximately between 30 and 400 Hz, our upper limits are probing below even the more stringent limit set by accretion onto the surface, thus searching a physically significant portion of parameter space. Furthermore, our results are probing the parameter space predicted by the models of Glampedakis & Suvorov (2021) up to $f_0 \sim 500$ Hz. Searching at these higher GW frequencies is important;

although the frequency of Sco X-1 is not known, the distribution of spin frequencies of the observed AMXPs appears to be bimodal (Patruno et al. 2017), with a "fast" population of pulsars, for which Patruno et al. (2017) make the hypothesis that GW emission may play a role, centered around $\nu_s \approx 550$ Hz (i.e., $f_0 \approx 1100$ Hz for triaxial emission), and a slower population centered around $\nu_s \approx 300$ Hz (i.e., $f_0 \approx 600$ Hz).

To give an illustration of how our results can constrain possible torque balance models, consider in more detail the simplest accretion model, i.e., that obtained by setting $r = R_*$ in Equation (3). In this case we may ask whether, by comparing the upper limits from our searches to the theoretical value for the $h_0$ in Equation (3), it is possible to put constraints on the physical parameters of the star for which the torque balance scenario is still viable. To answer this question, we consider two parameters: the mass of the star and the inclination angle $\iota$, for two EOSs taken from the CompOSE database (Typel et al. 2015; Oertel et al. 2017; Typel et al. 2022) both for a softer EOS, GR15 (Gulminelli & Raduta 2015), and for a stiffer EOS, GPPVA (Grill et al. 2014). The results can be seen in Figure 8 for the mass of the NSs. We see that, for the range of GW frequencies in which our search is most sensitive, we can exclude the torque balance scenario for higher-mass NSs, especially in the case of a stiffer EOS. The GW amplitude is, however, clearly very strongly affected by the inclination angle, so in Figure 8 we also plot our constraints in terms of the inclination angle $\iota$, holding the stellar mass fixed at $M = 1.4 M_\odot$. In this case we also see that we can rule out torque balance models with nearly circular polarization (small $\iota$) over a wide range of frequencies for both choices of EOS.

## 6. Conclusions and Outlook

We have presented the results of the most sensitive search to date for GWs from Sco X-1, using LIGO detector data from the third observing run of Advanced LIGO and Advanced Virgo. We have set upper limits across a range of GW signal frequencies $25\,\text{Hz} < f_0 < 1600\,\text{Hz}$, corresponding to NS spin frequencies of $12.5\,\text{Hz} < \nu_s < 800\,\text{Hz}$. The sensitivity of our search is now probing possible models of torque balance equilibrium over a range of GW frequencies spanning hundreds of hertz and, for the first time, approaches the standard conservative torque balance prediction even under pessimistic assumptions about NS inclination angle. We expect to see a further improvement in sensitivity from upcoming LIGO-Virgo-KAGRA observing runs (Abbott et al. 2020).

This material is based on work supported by NSF's LIGO Laboratory, which is a major facility fully funded by the National Science Foundation. The authors also gratefully acknowledge the support of the Science and Technology Facilities Council (STFC) of the United Kingdom, the Max-Planck-Society (MPS), and the State of Niedersachsen/Germany for support of the construction of Advanced LIGO and construction and operation of the GEO 600 detector. Additional support for Advanced LIGO was provided by the Australian Research Council. The authors gratefully acknowledge the Italian Istituto Nazionale di Fisica Nucleare (INFN), the French Centre National de la Recherche Scientifique (CNRS), and the Netherlands Organization for Scientific Research (NWO) for the construction and operation of the Virgo detector and the creation and support of the EGO consortium. The authors also gratefully acknowledge research support from these agencies, as well as by the Council of Scientific and Industrial Research of India, the Department of Science and Technology, India, the Science & Engineering Research Board (SERB), India, the Ministry of Human Resource Development, India, the Spanish Agencia Estatal de Investigación (AEI), the Spanish Ministerio de Ciencia e Innovación and Ministerio de Universidades, the Conselleria de Fons Europeus, Universitat i Cultura and the Direcció General de Política Universitaria i Recerca del Govern de les Illes Balears, the Conselleria d'Innovació Universitats, Ciència i Societat Digital de la Generalitat Valenciana and the CERCA Programme Generalitat de Catalunya, Spain, the National Science Centre of Poland and the European Union—European Regional Development Fund, Foundation for Polish Science (FNP), the Swiss National Science Foundation (SNSF), the Russian Foundation for Basic Research, the Russian Science Foundation, the European Commission, the European Social Funds (ESF), the European Regional Development Funds (ERDF), the Royal Society, the Scottish Funding Council, the Scottish Universities Physics Alliance, the Hungarian Scientific Research Fund (OTKA), the French Lyon Institute of Origins (LIO), the Belgian Fonds de la Recherche Scientifique (FRS-FNRS), Actions de Recherche Concertées (ARC) and Fonds Wetenschappelijk Onderzoek—Vlaanderen (FWO), Belgium, the Paris Île-de-France Region, the National Research, Development and Innovation Office Hungary (NKFIH), the National Research Foundation of Korea, the Natural Science and Engineering Research Council Canada, Canadian Foundation for Innovation (CFI), the Brazilian Ministry of Science, Technology, and Innovations, the International Center for Theoretical Physics South American Institute for Fundamental Research (ICTP-SAIFR), the Research Grants Council of Hong Kong, the National Natural Science Foundation of China (NSFC), the Leverhulme Trust, the Research Corporation, the Ministry of Science and Technology (MOST), Taiwan, the United States Department of Energy, and the Kavli Foundation. The authors gratefully acknowledge the support of the NSF, STFC, INFN, and CNRS for provision of computational resources.

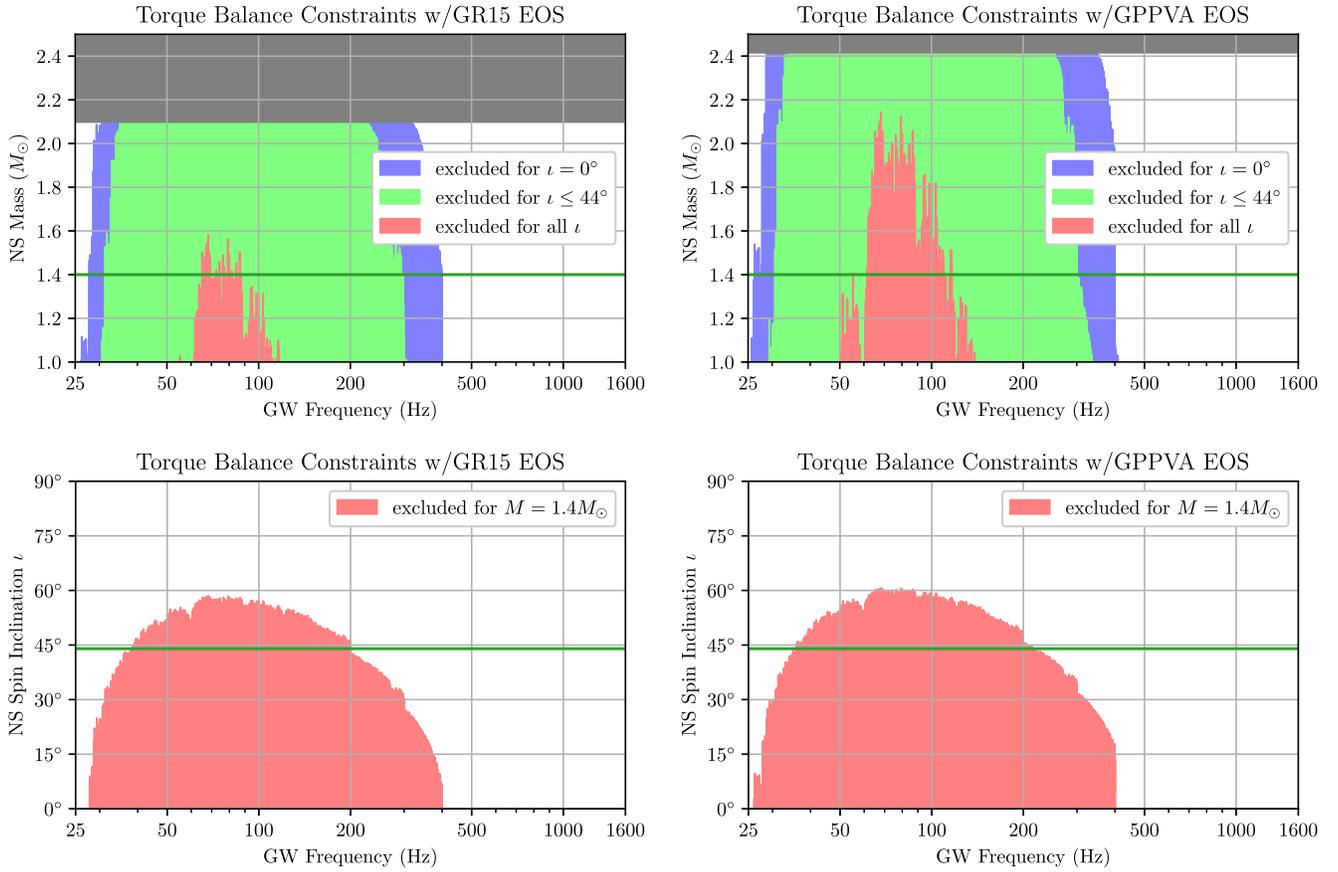

**Figure 8.** Illustration of how the upper limits shown in Figure 6 can constrain models for Sco X-1 that include torque balance due to gravitational waves. In the top row, we show constraints on the NS mass, assuming torque balance due to GW at the specified frequency, in the simple model where $r = R_*$. We consider two EOSs, a softer model, GR15 (Gulminelli & Raduta 2015), and a stiffer model, GPPVA (Grill et al. 2014). The largest exclusion region is for an NS inclination angle $\iota = 0°$ (or 180°), where the GWs would be circularly polarized. Assuming the worst-case scenario of linear polarization $\iota = 90°$ gives constraints that hold for any inclination, and the most likely value of $\iota = 44°$ (aligned with the binary orbit) gives an intermediate case. We see that for both EOSs (and especially for the stiffer EOS) the torque balance scenario can be excluded for higher-mass NSs for the GW frequency range in which our searches are more sensitive. Since the inclination angle plays a strong role in the constraints, we present in the bottom row, for the same EOSs but for a fixed mass of $M = 1.4 M_\odot$, the limits that can be set on $\iota$. We see that, for both EOSs, our observations can exclude nearly circular polarized (small $\iota$) GW emission at the torque balance level for a wide range of frequencies.

This work was supported by MEXT, JSPS Leading-edge Research Infrastructure Program, JSPS Grant-in-Aid for Specially Promoted Research 26000005, JSPS Grant-in-Aid for Scientific Research on Innovative Areas 2905: JP17H06358, JP17H06361, and JP17H06364, JSPS Core-to-Core Program A. Advanced Research Networks, JSPS Grant-in-Aid for Scientific Research (S) 17H06133 and 20H05639, JSPS Grant-in-Aid for Transformative Research Areas (A) 20A203: JP20H05854, the joint research program of the Institute for Cosmic Ray Research, University of Tokyo, National Research Foundation (NRF), Computing Infrastructure Project of KISTI-GSDC, Korea Astronomy and Space Science Institute (KASI), and Ministry of Science and ICT (MSIT) in Korea, Academia Sinica (AS), AS Grid Center (ASGC) and the Ministry of Science and Technology (MoST) in Taiwan under grants including AS-CDA-105-M06, Advanced Technology Center (ATC) of NAOJ, and Mechanical Engineering Center of KEK.

This paper has been assigned LIGO Document No. LIGO-P2100110-v13.

*Software:* LALSuite (LIGO Scientific Collaboration 2018), LatticeTiling (Wette 2014), PyFstat (Ashton & Prix 2018; Keitel et al. 2021; Ashton et al. 2022) numpy (Harris et al. 2020), matplotlib (Hunter 2007), scipy (Virtanen et al. 2020).